\documentclass[twocolumn]{aastex6}

\def\adfs27{ADFS-27}

\received{October 27, 2020}
\revised{November 27, 2020}
\accepted{November 29, 2020}

\usepackage{apjfonts}
\usepackage{aas_macros, xspace} 
\newcommand{\eq}{\,=\,}

\def\ts     {\thinspace}
\def\kms    {\ts km\ts s$^{-1}$}

\def\msol   {$M_{\odot}$}
\def\lsol   {$L_{\odot}$}

\def\aco    {{\rm CO}($J$=1$\to$0)}

\def\bco    {{\rm CO}($J$=2$\to$1)}

\def\eco    {{\rm CO}($J$=5$\to$4)}
\def\fco    {{\rm CO}($J$=6$\to$5)}

\def\hco    {{\rm CO}($J$=8$\to$7)}
\def\ico    {{\rm CO}($J$=9$\to$8)}
\def\jco    {{\rm CO}($J$=10$\to$9)}

\def\water   {{\rm H$_2$O}(2$_{11}$$\to$2$_{02}$)}
\def\waterb   {{\rm H$_2$O}(3$_{12}$$\to$2$_{21}$)}
\def\waterc   {{\rm H$_2$O}(3$_{21}$$\to$3$_{12}$)}

\shorttitle{Gas Excitation and Feedback in an Extremely Red Binary HyLIRG at $z$$\sim$6}
\shortauthors{Riechers et al.}

\begin{document}

\title{Rise of the Titans:\ Gas Excitation and Feedback in a Binary Hyper-Luminous Dusty Starburst Galaxy at $z$$\sim$6}


\author{Dominik A.\ Riechers\altaffilmark{1}}
\author{Hooshang Nayyeri\altaffilmark{2}}
\author{Denis Burgarella\altaffilmark{3}}
\author{Bjorn H.\ C.\ Emonts\altaffilmark{4}}
\author{\\ David L.\ Clements\altaffilmark{5}}
\author{Asantha Cooray\altaffilmark{2}}
\author{Rob J.\ Ivison\altaffilmark{6}}
\author{Seb Oliver\altaffilmark{7}}
\author{\\ Ismael P\'erez-Fournon\altaffilmark{8,9}}
\author{Dimitra Rigopoulou\altaffilmark{10}}
\author{Douglas Scott\altaffilmark{11}}    

\altaffiltext{1}{Cornell University, Space Sciences Building, Ithaca, NY 14853, USA}
\altaffiltext{2}{Department of Physics and Astronomy, University of California, Irvine, CA 92697, USA}
\altaffiltext{3}{Aix Marseille Univ, CNRS, CNES, LAM, Marseille, France}
\altaffiltext{4}{National Radio Astronomy Observatory, 520 Edgemont Road, Charlottesville, VA 22903, USA}
\altaffiltext{5}{Astrophysics Group, Imperial College London, Blackett Laboratory, Prince Consort Road, London SW7 2AZ, UK}
\altaffiltext{6}{European Southern Observatory, Karl-Schwarzschild-Stra{\ss}e 2, D-85748 Garching, Germany}
\altaffiltext{7}{Astronomy Centre, Department of Physics and Astronomy, University of Sussex, Brighton BN1 9QH, UK}
\altaffiltext{8}{Instituto de Astrofisica de Canarias, E-38200 La Laguna, Tenerife, Spain}
\altaffiltext{9}{Departamento de Astrofisica, Universidad de La Laguna, E-38205 La Laguna, Tenerife, Spain}
\altaffiltext{10}{Department of Physics, University of Oxford, Keble Road, Oxford OX1 3RH, UK}
\altaffiltext{11}{Department of Physics and Astronomy, University of British Columbia, 6224 Agricultural Road, Vancouver, BC V6T 1Z1, Canada}

 \email{riechers@cornell.edu}

\begin{abstract}

  We report new observations toward the hyper-luminous dusty
  starbursting major merger \adfs27\ ($z$=5.655), using the Australia
  Telescope Compact Array (ATCA) and the Atacama Large
  Millimeter/submillimeter Array (ALMA).  We detect \bco, \hco, \ico,
  \jco, and \waterb\ emission, and a P-Cygni-shaped
  OH$^+$(1$_1$$\to$0$_1$) absorption/emission feature.  We also
  tentatively detect \waterc\ and OH$^+$(1$_2$$\to$0$_1$) emission and
  CH$^+$($J$=1$\to$0) absorption.  We find a total cold molecular mass
  of $M_{\rm gas}$=(2.1$\pm$0.2)$\times$10$^{11}$\,$(\alpha_{\rm
    CO}/1.0)$\,\msol. We also find that the excitation of the
  star-forming gas is overall moderate for a $z$$>$5 dusty starburst,
  which is consistent with its moderate dust temperature.  A high
  density, high kinetic temperature gas component embedded in the gas
  reservoir is required to fully explain the CO line ladder.  This
  component is likely associated with the ``maximum starburst'' nuclei
  in the two merging galaxies, which are separated by only
  (140$\pm$13)\,\kms\ along the line of sight and 9.0\,kpc in
  projection. The kinematic structure of both components is consistent
  with galaxy disks, but this interpretation remains limited by the
  spatial resolution of the current data. The OH$^+$ features are only
  detected towards the northern component, which is also the one that
  is more enshrouded in dust and thus remains undetected up to
  1.6\,$\mu$m even in our sensitive new {\em Hubble Space Telescope}
  Wide Field Camera 3 imaging. The absorption component of the OH$^+$
  line is blueshifted and peaks near the CO and continuum emission
  peak while the emission is redshifted and peaks offset by 1.7\,kpc
  from the CO and continuum emission peak, suggesting that the gas is
  associated with a massive molecular outflow from the intensely
  star-forming nucleus that supplies 125\,\msol\,yr$^{-1}$ of enriched
  gas to its halo.

\end{abstract}


\section{Introduction} \label{sec:intro}

With infrared luminosities $L_{\rm IR}$$>$10$^{13}$\,\lsol,
hyper-luminous infrared galaxies (HyLIRGs; e.g., Sanders \& Mirabel
\citeyear{sm96}; Rowan-Robinson \citeyear{rowan00}) represent the most
intensely star-forming galaxies through cosmic history. While rare at
any epoch, they are particularly exceptional in the early universe
within the first billion years, when structure formation had not yet
sufficiently matured to allow for the presence of a substantial
abundance of the most massive dark matter halos that are thought to
host such systems (e.g., Robson et al.\ \citeyear{robson14}, and
references therein). The majority of their star formation activity,
taking place at rates of SFR$>$1000\,\msol\,yr$^{-1}$, and even the
already existing stellar populations, are commonly hidden from our
view due to high levels of dust obscuration (e.g., Simpson et
al.\ \citeyear{simpson20}, and references therein). This makes them
the most luminous, and perhaps most massive tail of the dusty
star-forming galaxy (DSFG) population (e.g., Hodge \& da Cunha
\citeyear{hd20}, and references therein). The star formation activity
in HyLIRGs is typically driven by major mergers (e.g., Engel et
al.\ \citeyear{engel10}; Oteo et al.\ \citeyear{oteo16}; Riechers et
al.\ \citeyear{riechers17}). It commonly takes place in compact
regions only one to a few kiloparsecs across, which are permeated by
intense radiation fields powered by the large quantities of massive
stars that were just born, and impacted by mechanical feedback due to
ejecta from evolved stars and supernova explosions (e.g., Riechers et
al.\ \citeyear{riechers13b,riechers14b,riechers17,riechers20a}; Fu et
al.\ \citeyear{fu13}; Ivison et al.\ \citeyear{ivison13}; Tadaki et
al.\ \citeyear{tadaki20}). This radiation pressure and mechanical
feedback can lead to large-scale turbulence, winds and outflows, which
deposit enriched gas in the galaxy's halo. This gas can be traced
through absorption in molecular species like OH$^+$ and CH$^+$, back
to the first billion years of cosmic history, as demonstrated by
detections in the $z$$>$6 HyLIRG HFLS3 (Riechers et
al.\ \citeyear{riechers13b}) and detailed investigations of the line
profiles in a sample of strongly-lensed $z$$\sim$2 DSFGs (Falgarone et
al.\ \citeyear{falgarone17}; Indriolo et al.\ \citeyear{indriolo18}).
These compact starburst regions are commonly characterized by a high
excitation of the molecular gas due to a high gas density and kinetic
temperature (as traced by the CO rotational line ladder), and they may
be embedded in more extended, massive cold gas reservoirs that supply
the material to support the starburst activity on tens to 100\,Myr
timescales (e.g., Greve et al.\ \citeyear{greve05}; Tacconi et
al.\ \citeyear{tacconi08}; Riechers et
al.\ \citeyear{riechers10a,riechers11e,riechers20a}; Ivison et
al.\ \citeyear{ivison11}; Hodge et al.\ \citeyear{hodge15}).

We here report new observations of the hyper-luminous binary starburst
\adfs27\ at $z$=5.655 (2HERMES S250 SF J043657.7--543810; Riechers et
al.\ \citeyear{riechers17}) with ATCA, ALMA, and the {\em Hubble Space
  Telescope}, to obtain a more detailed understanding of the physical
properties of such exceptionally luminous systems. ADFS-27 was
initially discovered in the {\em Herschel} Multi-tiered Extragalactic
Survey (HerMES; Oliver et al.\ \citeyear{oliver12}) combined with
follow-up observations with the Atacama Pathfinder EXperiment (APEX)
telescope through its exceptionally ``red'' color between the 250 and
870\,$\mu$m bands, which make it the only known dusty point source in
the survey that fulfills the ``870\,$\mu$m riser'' criterion (i.e.,
$S_{250\,\mu{\rm m}} < S_{350\,\mu{\rm m}} < S_{500\,\mu{\rm m}} <
S_{870\,\mu{\rm m}}$; Riechers et al.\ \citeyear{riechers17}). This
work investigates the stellar light emerging from the system, the mass
of its gas reservoir, the gas excitation, and evidence for
starburst-driven feedback. We present the data and their calibration
in Section 2, before discussing the immediate results and presenting a
broader analysis, informed by models, in Sections 3 and 4. A summary and
conclusions are given in Section 5. We use a concordance, flat
$\Lambda$CDM cosmology throughout, with
$H_0$\eq69.6\,\kms\,Mpc$^{-1}$, $\Omega_{\rm M}$\eq0.286, and
$\Omega_{\Lambda}$\eq0.714. At $z$=5.655, 1$''$ on the sky corresponds
to 6.026\,kpc with these parameters, the luminosity distance is
$D_{\rm L}$=55044.8\,Mpc, and the time since the Big Bang is $t_{\rm
  cos}$=1.016\,Gyr.

\section{Data} \label{sec:data}

\subsection{ATCA}

We observed the \bco\ line ($\nu_{\rm rest}$=230.5380\,GHz, redshifted
to $\nu_{\rm obs}$=34.6413\,GHz) toward \adfs27\ using ATCA (project
ID: C3226; PI:\ Riechers). Observations were carried out under
acceptable to good weather conditions for two tracks on 2017 October
05 and 06 using the 7\,mm band receivers in the H168 configuration
(baseline range:\ 61--185\,m for the 5-antenna main array, with a
sixth outrigger antenna providing 4.3--4.5\,km baselines), totaling
16\,hr of observing time. The outrigger antenna was only available for
the first track, and it was manually flagged due to the significantly
poorer phase stability on the longest baselines. The nearby bright
radio quasar 0454--463 ($\sim$1.4\,Jy) at $\sim$9\,deg distance from
\adfs27\ was observed for phase and bandpass calibration every
10\,minutes. Pointing corrections were obtained on a calibrator
source roughly every hour, or when moving the telescopes by $>$20\,deg
on the sky.  Absolute fluxes were determined based on observations of
PKS\,1934--628, which we cross-checked based on observations of
Uranus. From our data, we estimate that the absolute flux scale is
reliable at the 12\%--20\% level, with a relative accuracy at the 2\%
level between the two tracks.

The correlator was set up using the Compact Array Broadband Backend
(CABB; Wilson et al.\ \citeyear{wilson11}) in CFB\,1M mode with a
bandwidth of 2\,GHz in each of two intermediate frequency (IF) band at
a spectral resolution of 1\,MHz (8.7\,\kms ). Data were binned to
40--80\,\kms\ resolution in the subsequent analysis.

Data reduction was performed using the {\sc miriad} package (Sault et
al.\ \citeyear{sault95}). Imaging the line emission yields a
synthesized beam size of 7\farcs2$\times$5\farcs3 at an rms noise
level of 54.3\,$\mu$Jy\,beam$^{-1}$ over 1509\,\kms. Imaging the data
across the entire line-free bandwidth yields a continuum rms level of
12\,$\mu$Jy\,beam$^{-1}$.


\begin{figure*}
\begin{deluxetable}{ l c c c c c c }
\tablecaption{\adfs27 imaging parameters \label{t0}}
\tablehead{
Line or $\lambda_{\rm cont}$ & configuration & weighting & beam size & rms noise & frequency range & velocity range \\
         & (ATCA/ALMA) &           & $\theta_{\rm maj}$$\times$$\theta_{\rm min}$ & ($\mu$Jy\,beam$^{-1}$) & & (\kms ) }
\startdata
\bco & H168          & natural   & 7\farcs2$\times$5\farcs3   & 54.3 & 175\,MHz & 1509 \\
\hco & C43-1/2       & natural   & 3\farcs1$\times$2\farcs3   & 100  & 400\,MHz & 866 \\
\ico & C43-5         & natural   & 0\farcs56$\times$0\farcs49 & 23.8 & 593.75\,MHz & 1142 \\
     &               & robust +0.5 & 0\farcs42$\times$0\farcs36 & 28 & & \\
\jco & C43-2/3+5     & natural   & 0\farcs54$\times$0\farcs45 & 35   & 750\,MHz & 1299 \\
     &               & robust +0.5 & 0\farcs40$\times$0\farcs34 & 40 & & \\
CH$^+$($J$=1$\to$0) abs.\ & C43-1/2 & natural & 3\farcs3$\times$2\farcs6 & 130 & 460.938\,MHz & 1100 \\
OH$^+$(1$_1$$\to$0$_1$) abs.\ & C43-5 & natural & 0\farcs56$\times$0\farcs49 & 29 & 343.75\,MHz & 664 \\
OH$^+$(1$_1$$\to$0$_1$) em.\ & C43-5 & natural & 0\farcs56$\times$0\farcs49 & 28 & 421.875\,MHz & 815 \\
OH$^+$(1$_2$$\to$0$_1$) em.\ & C43-5 & natural & 0\farcs60$\times$0\farcs52 & 42 & 156.25\,MHz & 321 \\
2.3\,mm continuum & C43-1/2   & natural      & 3\farcs3$\times$2\farcs6 & 34 & 3.43\,GHz & \\
2.2\,mm continuum & C43-1/2   & natural      & 3\farcs0$\times$2\farcs3 & 36 & 3.12\,GHz & \\
2.0\,mm continuum & C43-5     & robust --0.5 & 0\farcs33$\times$0\farcs28 & 20 & 3.43\,GHz & \\
1.9\,mm continuum & C43-5     & robust --0.5 & 0\farcs32$\times$0\farcs26 & 29 & 1.74\,GHz & \\
1.7\,mm continuum & C43-2/3+5 & robust --0.5 & 0\farcs31$\times$0\farcs28 & 50 & 2.08/0.92\,GHz\tablenotemark{a} & \\
\enddata 
\tablenotetext{\rm a}{Line-free frequency ranges for compact/extended configuration observations, respectively.}
\end{deluxetable}
\vspace{-10mm}
\end{figure*}


\subsection{ALMA}

We observed the \hco, \ico, and \jco\ lines ($\nu_{\rm
  rest}$=921.7997, 1036.9124, and 1151.9855\,GHz; redshifted to
$\nu_{\rm obs}$=138.5124, 155.8095, and 173.1007\,GHz, respectively)
in setups that also covered the CH$^+$($J$=1$\to$0),
OH$^+$(1$_1$$\to$0$_1$), H$_2$O(3$_{12}$$\to$2$_{21}$), and
H$_2$O(3$_{21}$$\to$3$_{12}$) lines ($\nu_{\rm rest}$=835.0790,
1033.0582, 1153.1268, and 1162.9116\,GHz; $\nu_{\rm obs}$=125.4814,
155.2304, 173.2723, and 174.7425\,GHz, respectively) toward
\adfs27\ using ALMA (project ID:\ 2017.1.00235.S; PI:\ Riechers; we
also included archival data from project 2018.1.00966.S in the
frequency range overlapping with our science goals). Observations were
carried out under acceptable to very good weather conditions during
six runs in cycles 5 and 6 between 2018 July 14 and November 20 using
the band 4 and 5 receivers on 42--48 12\,m antennas in a compact array
configuration for \hco\ (C43-1/2; baseline range:\ 15--313\,m), in a
moderately extended configuration for \ico\ (C43-5;
15\,m--1.4\,km),\footnote{Observations in a more compact configuration
  were scheduled as part of project 2017.1.00235.S, but not carried
  out.} and in both a compact and a moderately extended configuration
for \jco\ (C43-2/3 and C43-5; 15\,m--1.4\,km). A total of 12.6, 76,
and 102\,minutes on source were spent for observing the \hco, \ico,
and \jco\ lines, respectively. The nearby radio quasars J0425--5331 or
J0441--5154 were observed regularly for phase calibration. J0519--4546
was used for pointing, amplitude, bandpass and absolute flux
calibration, leading to $<$10\% calibration uncertainty.

The correlator was set up with two spectral windows of 1.875\,GHz
bandwidth (dual polarization) each per sideband, at a sideband
separation of 8\,GHz for all \hco\ and \ico\ observations, and the
\jco\ observations in a compact configuration. The upper sideband for
the \jco\ observations (which did not contain any lines of interest)
fell into the outer wing of the 183\,GHz atmospheric H$_2$O line, and
had to be discarded due to unreliable calibration given the modest
weather conditions at the time of observing. For the extended
configuration \jco\ observations, three partially overlapping spectral
windows were placed in the lower sideband (only one of which was used
in the analysis presented here), and none was placed in the upper
sideband.

Data reduction was performed using version 5.4.0 or 5.6.1 of the {\sc
  casa} package (McMullin et al.\ \citeyear{mcmullin07}), aided by the
calibration pipeline included with each version. Data were mapped
manually using the CLEAN algorithm via the {\tt tclean} task with
``natural'', Briggs robust +0.5, and Briggs robust --0.5 weighting,
which resulted in the image parameters detailed in
Table~\ref{t0}. Continuum maps are created over the entire line-free
bandwidths.

\begin{figure*}[tbh]
\epsscale{1.18}
\plotone{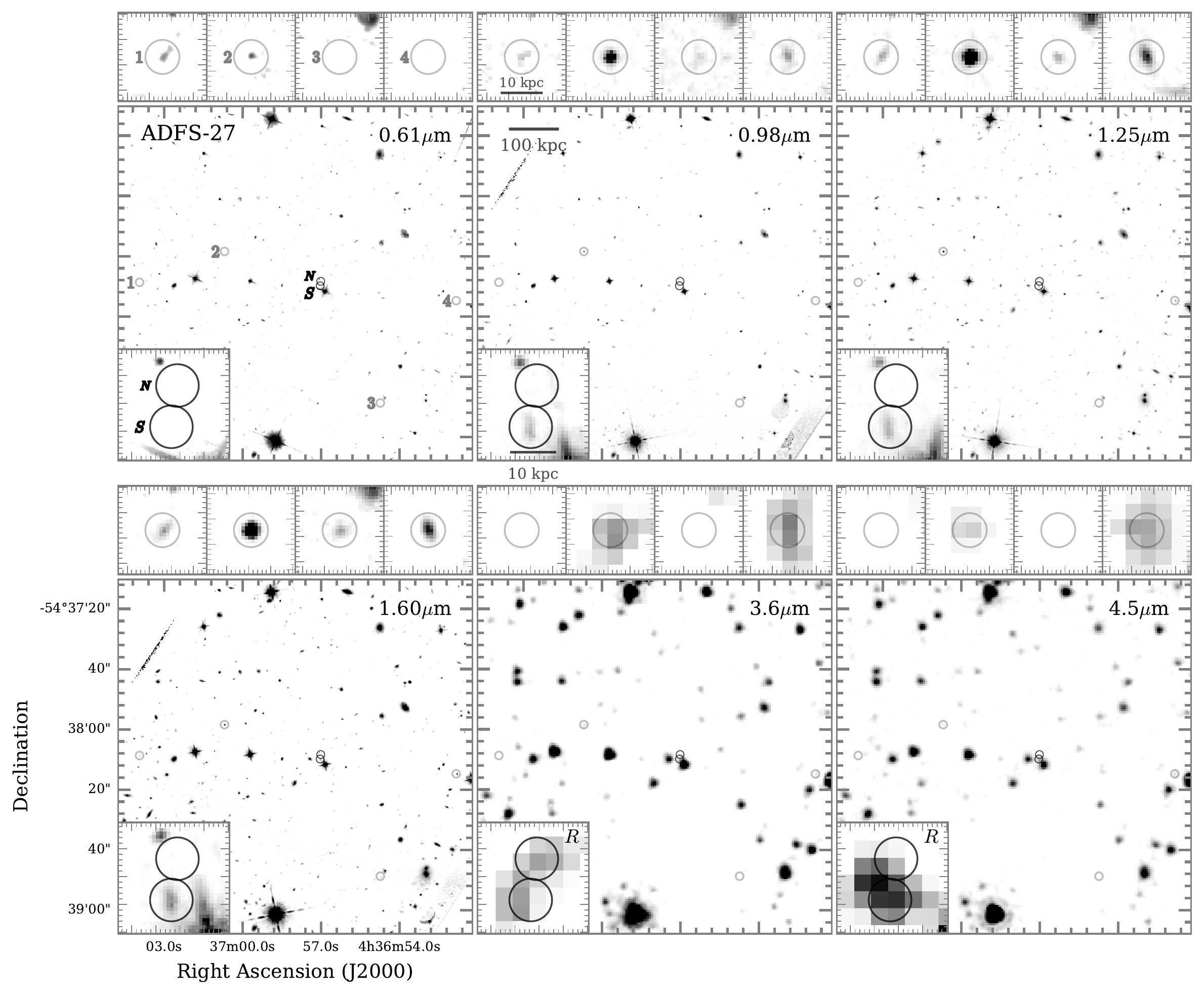}
\vspace{-5mm}

\caption{{\em Hubble Space Telescope} ACS/F606W and WFC3-IR F098M,
  F125W, and F160W, and {\em Spitzer} IRAC ch1 and ch2 imaging (left
  to right and top to bottom; insets are zoomed-in and show residuals after de-blending
  from foreground sources in the IRAC bands where applicable) toward
  the field around \adfs27\ ({\em Spitzer} data adopted from Riechers
  et al.\ \citeyear{riechers17}). Black circles centered on the
  870\,$\mu$m continuum positions (Riechers et
  al.\ \citeyear{riechers17}) of ADFS-27N and S are shown for
  reference. ADFS-27N is not detected up to at least
  1.6\,$\mu$m. ADFS-27S is not detected in the F606W band (rest-frame
  910\,nm), but in all other bands (rest-frame 1500--6800\,nm). No
  candidate lensing galaxy is detected, which is consistent with both
  source components not being strongly gravitationally lensed, but
  instead intrinsically very luminous. Gray circles and panels on top
  (same zoom factors) show the positions of candidate Lyman-break
  galaxies at the redshift of \adfs27.\label{f0}}
%
\end{figure*}

\subsection{Hubble Space Telescope}

\adfs27\ was observed with the {\em Hubble Space Telescope} Advanced
Camera for Surveys Wide-Field Camera (ACS-WFC) in the F606W filter
(effective wavelength:\ 581\,nm) for 2 orbits on 2020 August 11/12 and
with the Wide-Field Camera 3 in the infrared F098M, F125W, and F160W
filters (WFC3-IR; 983, 1236, and 1528\,nm) for 1, 1, and 2 orbits on
2018 August 18 and 2020 June 14, respectively (program ID:\ 15919,
PI:\ Riechers for all except F125W, which is from 15464,
PI:\ Long). The total exposure times in the F606W, F098M, F125W, and
F160W filters were 4800, 2712, 2612, and 5612\,s, respectively

Standard procedures were used to calibrate the data, including flat
fielding and flux calibration, as well as masking of bad pixels, bias
and dark current subtractions, corrections for geometric distortions,
and the combination of exposures with {\tt AstroDrizzle}. Initial
catalogs for astrometry were obtained by running {\sc SExtractor}
(Bertin \& Arnouts \citeyear{ba96}) in single mode for each
band. Astrometry was carried out by matching the images in each filter
to the GAIA DR2 catalog (Gaia Collaboration \citeyear{gaia18}) with a
matching radius of one full width at half maximum, which resulted in
typically 14 matches per band in the WFC3 filters. The final shift
applied was taken as the mean of the shifts for these matched
sources. Photometry catalogs were made using {\sc SExtractor} in dual
mode, with F160W as the detection image, and using the other bands
iteratively to obtain the photometry.

\section{Results}

\subsection{Continuum Emission}

We here report our new findings on the rest-frame ultraviolet to
millimeter continuum emission toward \adfs27. The main results are
summarized in Table~\ref{t1}.

\subsubsection{HST}

The F606W filters sits below the Lyman break at $z$$\sim$5.7. Thus, no
emission is detected toward \adfs27, as expected
(Fig.~\ref{f0}). Also, there is no evidence for the presence of a
foreground galaxy that could cause strong gravitational lensing. The
nearest bright foreground galaxy detected in the WFC3-IR filters is
3.7$''$ to the east of ADFS-27S. There is a faint object at 1.0$''$
distance from ADFS-27N detected in F606W, but unlike most effective
deflectors, it is not a massive red galaxy, such that strong lensing
(i.e., a flux magnification factor of $\mu_{\rm L}$$\geq$2) remains
unlikely. A small amount of magnification (a few per cent to few tens
of per cent) remains possible, but there is no evidence for a
significant extension of ADFS-27N along the axis that would be
expected for distortion due to weak lensing magnification (which lies
close to its minor axis). Also, there is no evidence for a foreground
galaxy group or cluster that could provide a larger-scale lensing
potential. As such, we confirm that ADFS-27 is intrinsically a
HyLIRG. ADFS-27N also remains undetected in all WFC3-IR filters, while
spatially-resolved stellar emission is detected towards ADFS-27S in
all three filters (Fig.~\ref{f0}). This suggests that the stellar
light in ADFS-27N is significantly more heavily dust-obscured than in
ADFS-27S, such that no direct stellar mass measurement is possible for
the northern galaxy without the investment of at least many tens of
orbits of {\em HST} time, at least until the launch of the {\em James
  Webb Space Telescope} ({\em JWST}). Aperture fluxes and upper limits
are reported in Table~\ref{t1}.

We have also conducted a $V$-band dropout search for candidate
Lyman-break galaxies (LBGs) at the same redshift as \adfs27. We
initially selected the F606W and F098M filters, because the Lyman
break falls between these filters at $z$=5.655. We used selection
criteria defined in analogy with those used by Bouwens et
al.\ (\citeyear{bouwens15}), but fine-tuned toward the selection of
LBGs at $z$=5.6 (R.\ Bouwens 2020, private communication). First, we
required a detection at $>$5$\sigma$ in all WFC3-IR bands. Then, we
applied the following color selection criteria (in AB magnitudes):\\

\indent F606W -- F098M $>$ 2.3 \\
\indent F606W -- F098M $>$ 2.3 + (F098M -- F160W)\\
\indent F098M -- F160W $<$ 1.2 \\

As such, we used the F606W--F098M color as the main selection
criterion to reject $z$$<$5 galaxies, and the slope between the
near-infrared bands to select against excessively steep spectral
energy distribution shapes above the expected Lyman break. We also
rejected stars by excluding point sources from the resulting
sample. Due to the faint selection thresholds required, we also
rejected sources near the edges of the WFC3 coverage, where a higher
incidence of spurious faint sources is expected. This results in four
candidates, A27.LBG-1 to 4 (labeled 1 to 4 in Fig.~\ref{f0}). Despite
the break selection criterion, A27.LBG-1 and 2 clearly show faint
detections in the F606W filter, and A27.LBG-2 shows a morphology that
resembles a point source. These sources thus are discarded from the
LBG candidate sample. None of the two surviving candidates appear to
have detections in the F606W filter. A27.LBG-3 is lacking a detection
in the {\em Spitzer} IRAC bands at 3.6 and 4.5\,$\mu$m, but it also is
the far faintest source in the WFC3 bands. We thus retain it in the
LBG candidate sample. A27.LBG-4 has strong detections in the IRAC
bands, and thus, is the strongest candidate. A27.LBG-3 and 4 are at
43.7 and 45.4$''$ (corresponding to 264 and 274\,kpc at $z$=5.655)
projected distance from ADFS-27S, respectively. Given the custom
selection criteria, narrow-band Lyman--$\alpha$ imaging and/or
follow-up spectroscopy is required to assess whether or not any of the
candidates is at the redshift of \adfs27. If confirmed, our findings
would be consistent with an overdensity of star-forming galaxies
within 300\,kpc around \adfs27.

To further quantify the density contrast compared to the field, we
calculated the field density of $z$=5.7 LBGs in the same manner as
done by Pavesi et al.\ (\citeyear{pavesi18a}), which is based on the
method developed by Smol\v{c}i\'c et
al.\ (\citeyear{smolcic17}). Adopting a redshift slice of
$\Delta{z}$=0.64 and using the photometric redshifts reported in the
COSMOS2015 catalog (Laigle et al.\ \citeyear{laigle16}), we find an
expected density in the field of $\Sigma_{\rm
  field}$$\sim$0.19\,arcmin$^{-2}$. We then calculate the overdensity
parameter relative to the field as $\delta_{\rm
  g}(r)$=$\frac{N_r}{\Sigma_{\rm field}\,\pi\,r^2}$--1. This yields an
estimated overdensity of $\delta_{\rm g}(r)$$\simeq$8--11 out to the
distance of A27.LBG-4, and $\simeq$6--9 within a 300\,kpc radius,
where the range indicate the difference between considering
\adfs27\ as a single source or two sources. Out to the same distances,
the overdensity around the $z$=5.7 galaxy CRLE studied by Pavesi et
al.\ (\citeyear{pavesi18a}) is $\delta_{\rm g}(r)$$\simeq$5 or 4,
respectively. At face value, the overdensity around \adfs27\ thus may
be more significant, but the relative uncertainties are dominated by
small number statistics.


\begin{figure*}
\begin{deluxetable}{ l c c c l c }
\tablecaption{\adfs27 continuum photometry \label{t1}}
\tablehead{
Wavelength & Flux density\tablenotemark{a} & & & Telescope & Reference \\
           & total                         & north & south  & & \\
 ($\mu$m) & (mJy) & (mJy) & (mJy) & & }
\startdata
0.606\tablenotemark{b} & $<$0.352$\times$10$^{-3}$ & $<$0.043$\times$10$^{-3}$ & $<$0.352$\times$10$^{-3}$ & {\em HST}/ACS-WFC & 1 \\
0.98\tablenotemark{b}  & (1.104$^{+0.091}_{-0.084}$)$\times$10$^{-3}$ & $<$0.26$\times$10$^{-3}$ & (1.104$^{+0.091}_{-0.084}$)$\times$10$^{-3}$ & {\em HST}/WFC3-IR & 1 \\
1.25\tablenotemark{b}  & (1.652$\pm$0.040)$\times$10$^{-3}$ & $<$0.12$\times$10$^{-3}$ & (1.652$\pm$0.040)$\times$10$^{-3}$ & {\em HST}/WFC3-IR & 1 \\
1.25                   & $<$0.015          & & & VISTA/VHS & 2 \\
1.60\tablenotemark{b}  & (1.599$\pm$0.046)$\times$10$^{-3}$ & $<$0.14$\times$10$^{-3}$ & (1.599$\pm$0.046)$\times$10$^{-3}$ & {\em HST}/WFC3-IR & 1 \\
1.65                   & $<$0.022          & & & VISTA/VHS & 2 \\
2.15                   & $<$0.020          & & & VISTA/VHS & 2 \\
3.6\tablenotemark{c}   & (2.33$\pm$0.74)$\times$10$^{-3}$ & & & {\em Spitzer}/IRAC & 2 \\
4.5\tablenotemark{c}   & (4.20$\pm$0.82)$\times$10$^{-3}$ & & & {\em Spitzer}/IRAC & 2 \\
12                     & $<$0.6          & & & {\em WISE} & 2 \\
22                     & $<$3.6          & & & {\em WISE} & 2 \\
110                    & $<$30           & & & {\em Herschel}/PACS & 2 \\
160                    & $<$57           & & & {\em Herschel}/PACS & 2 \\
250\tablenotemark{d}   & 14.3$\pm$2.3    & & & {\em Herschel}/SPIRE & 2 \\
350\tablenotemark{d}   & 19.1$\pm$2.3    & & & {\em Herschel}/SPIRE & 2 \\
500\tablenotemark{d}   & 24.0$\pm$2.7    & & & {\em Herschel}/SPIRE & 2 \\
870                    & 25.4$\pm$1.8    & & & APEX/LABOCA & 2 \\
870                    & 28.1$\pm$0.9    & 15.70$\pm$0.76 & 12.43 $\pm$ 0.56 & ALMA & 2 \\
1733                   & 3.73$\pm$0.09   & 2.37$\pm$0.05 & 1.36$\pm$0.08 & ALMA & 1 \\
1910                   & 2.67$\pm$0.05   & 1.71$\pm$0.05 & 0.96$\pm$0.02 & ALMA & 1 \\
2053                   & 2.08$\pm$0.05   & 1.34$\pm$0.04 & 0.74$\pm$0.03 & ALMA & 1 \\
2173                   & 1.84$\pm$0.06   & & & ALMA & 1 \\
2323                   & 1.37$\pm$0.06   & & & ALMA & 1 \\
3000                   & 0.512$\pm$0.023 & & & ALMA (scan) & 2 \\
8653                   & $<$0.036        & & & ATCA & 1 \\
\enddata 
\tablereferences{[1] This work; [2] Riechers et al.\ (\citeyear{riechers17}).}
\tablenotetext{\rm a}{Limits are 3$\sigma$.}
\tablenotetext{\rm b}{Fluxes extracted from a region where emission is seen in the F160W filter. Emission is only detected toward ADFS-27S. Upper limits for ADFS-27N are taken as 3$\sigma$ limits in an aperture that is the same size as used for ADFS-27S, but are not added to the total flux since no emission is seen in the images.}
\tablenotetext{\rm c}{Values obtained after de-blending from foreground sources. Flux is likely dominated by ADFS-27S.}
\tablenotetext{\rm d}{Uncertainties do not account for confusion noise, which
  is 5.9, 6.3, and 6.8\,mJy (1$\sigma$) at 250, 350, and 500\,$\mu$m,
  respectively \citep{nguyen10}.}
\end{deluxetable}
\vspace{-10mm}
\end{figure*}


\subsubsection{ALMA and ATCA}

\begin{figure*}
\epsscale{1.15}
\plotone{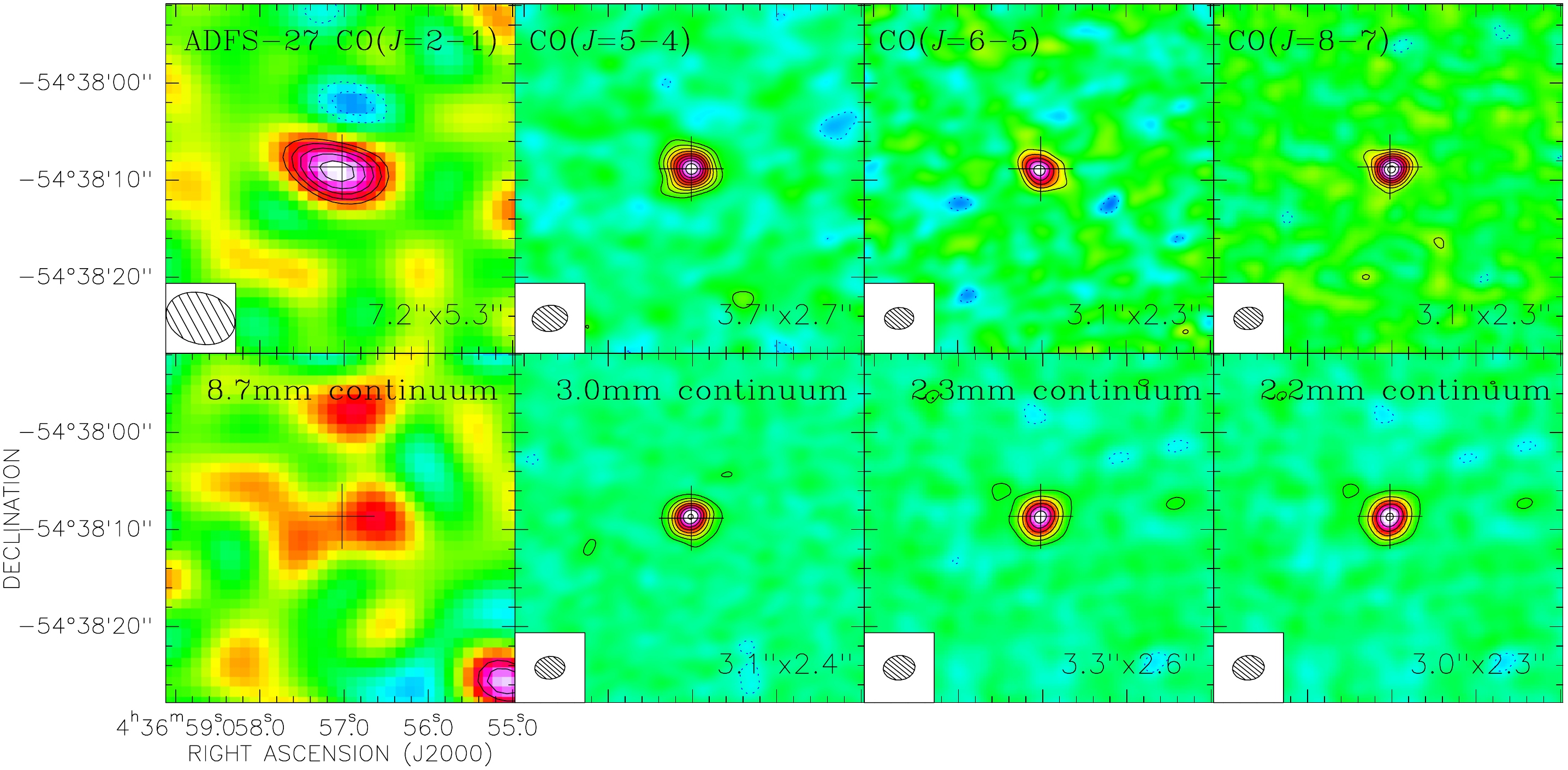}
\vspace{-2mm}

\caption{Maps of the CO $J$=2$\to$1 and continuum-subtracted CO
  $J$=5$\to$4, 6$\to$5, and 8$\to$7 line emission (top panels, left to
  right), and the 8.7 (not detected), 3.0, 2.3, and 2.2\,mm continuum
  emission (bottom panels) toward \adfs27\ (3\,mm data adopted from
  Riechers et al.\ \citeyear{riechers17}). Line contours are shown in
  steps of 1$\sigma$, 2$\sigma$, 2$\sigma$, and 2$\sigma$ (where
  1$\sigma$=0.082, 0.084, 0.12, and 0.087\,Jy\,\kms\,beam$^{-1}$ over
  1509, 651, 711, and 866\,\kms ), respectively, starting at
  $\pm$3$\sigma$. Continuum contours are shown in steps of 1$\sigma$,
  5$\sigma$, 5$\sigma$, and 5$\sigma$ (where 1$\sigma$=12, 11.2, 34,
  and 36\,$\mu$Jy\,beam$^{-1}$), respectively, starting at
  $\pm$3$\sigma$. The synthesized clean beam size is indicated in the
  bottom left corner of each panel where emission is detected. The
  cross in each panel indicates the peak position of the
  \eco\ emission. \label{f2}}
%
\end{figure*}

\begin{figure*}
\epsscale{1.15}
\plotone{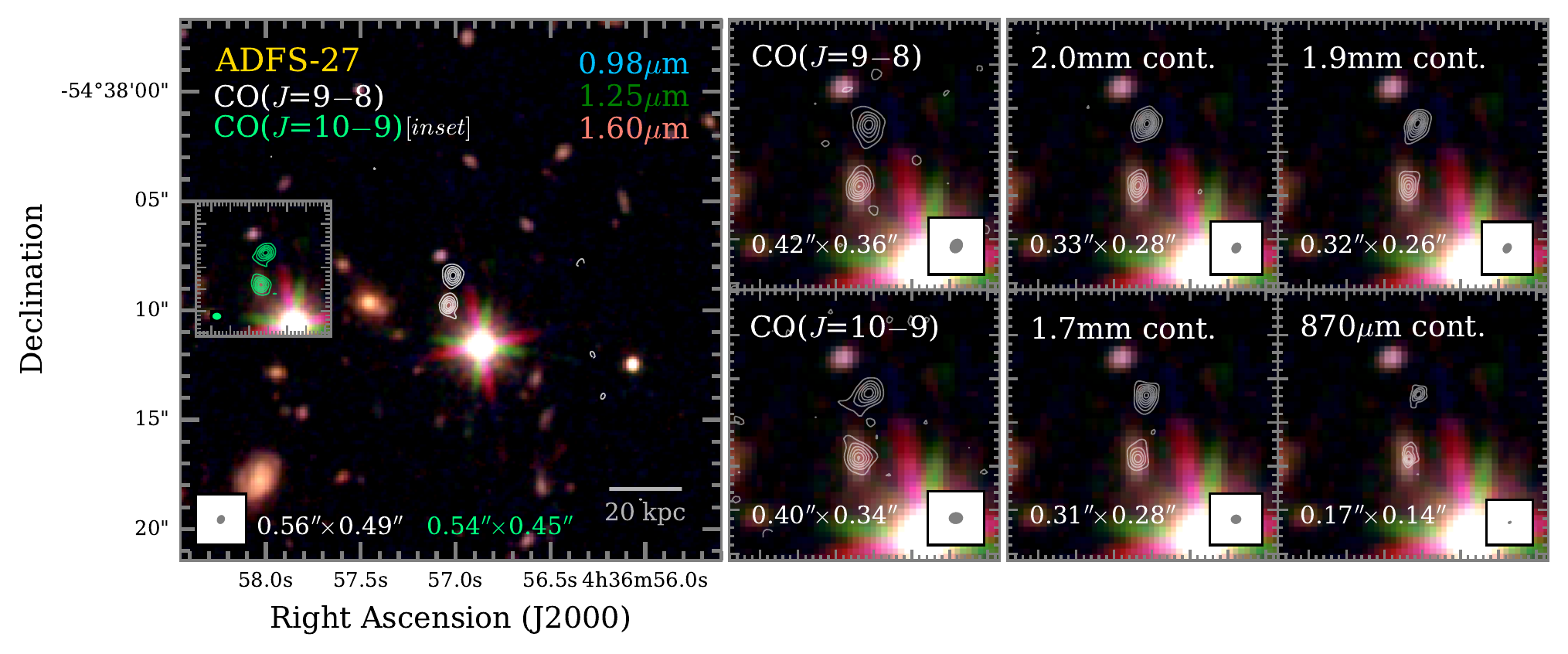}
\vspace{-2mm}

\caption{Maps of the CO $J$=9$\to$8 and 10$\to$9 (left panel and inset
  and middle panels), and 2.0, 1.9, 1.7, and 0.87\,mm continuum
  emission (right panels) toward \adfs27, overlaid on a color
  composite image of the 0.98, 1.25, and 1.60\,$\mu$m continuum
  emission. The 0.87\,mm data were adopted from Riechers et
  al.\ (\citeyear{riechers17}). Beam sizes are indicated in the bottom
  left or right corners of each panel. The left panel shows the CO
  $J$=9$\to$8 and 10$\to$9 (inset and light green labels) emission
  imaged over 1142 and 1299\,\kms\ with ``natural'' weighting. The
  middle panels shows the same, but imaged with Briggs robust 0.5
  weighting. Continuum emission is imaged with robust --0.5 weighting,
  except for 0.87\,mm, which is imaged with robust 0.5 weighting. Line
  contours are shown in steps of $\pm$4$\sigma$, $\pm$4$\sigma$
  (left), $\pm$3$\sigma$, and $\pm$3$\sigma$ (middle; where
  1$\sigma$=0.027, 0.045, 0.032, and 0.052\,Jy\,\kms\,beam$^{-1}$),
  respectively. Continuum contours are shown in steps of
  $\pm$4$\sigma$ (where 1$\sigma$=20, 29, 50, and
  108\,$\mu$Jy\,beam$^{-1}$).
  \label{f3}}
%
\end{figure*}

No continuum emission is detected toward \adfs27\ in the ATCA data,
and we report an upper limit in Table \ref{t1}. Continuum emission is
detected in all ALMA data between 1.7 and 2.3\,mm (Fig.~\ref{f2}), but
ADFS-27N and S are resolved apart from each other only shortwards of
2.1\,mm (Fig.~\ref{f3}) due to the more compact configuration chosen
for the \hco\ observations. The peak significance levels of the
detections are 31$\sigma$ and 42$\sigma$ at 2.3 and 2.2\,mm,
respectively. The peak significance levels of the detections of
ADFS-27N and S in maps imaged with natural weighting are 121$\sigma$
and 70$\sigma$ at 2.0\,mm, 90$\sigma$ and 56$\sigma$ at 1.9\,mm, and
89$\sigma$ and 56$\sigma$ at 1.7\,mm, respectively. Between 2.0\,mm
and 0.87\,mm, the continuum flux ratio between ADFS-27N and S
decreases from 1.81$\pm$0.09 to 1.26$\pm$0.08. This suggests that
ADFS-27S has an intrinsically ``warmer'' dust spectral energy
distribution shape than ADFS-27N. All continuum fluxes are summarized
in Table~\ref{t1}.

\begin{figure*}
\epsscale{1.15}
\plotone{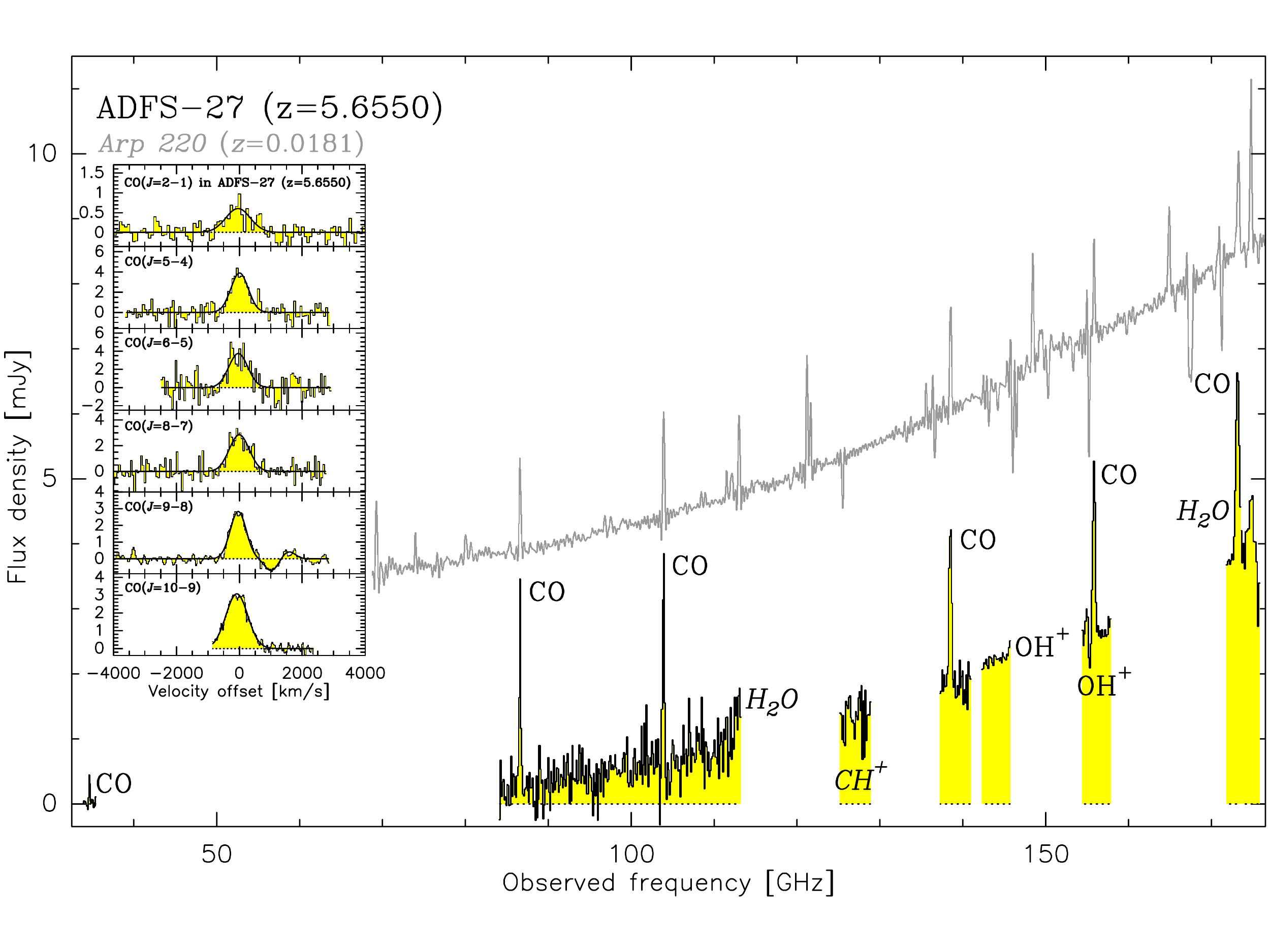}
\vspace{-9mm}

\caption{ALMA and ATCA broad-band spectrum of the line and continuum
  emission toward \adfs27\ (main panel), and CO spectral line profiles
  (inset). The 3\,mm data are adopted from Riechers et
  al.\ (\citeyear{riechers17}). The {\em Herschel}/SPIRE spectrum of
  the nearby ultra-luminous infrared galaxy Arp\,220 (gray; Rangwala
  et al.\ \citeyear{rangwala11}), shifted to the observed frame of
  \adfs27, is shown for comparison. The $\sim$9\,mm, 3--2.2\,mm, and
  $<$2.1\,mm data (histogram) in the main panel are shown at spectral
  resolutions of 77.5, 117.3, and 125\,MHz, respectively, except for
  the blue end of the spectrum, which is shown at 156\,MHz resolution
  due to higher noise. The CO $J$=2$\to$1 and continuum-subtracted CO
  $J$=5$\to$4, 6$\to$5, 8$\to$7, 9$\to$8 and 10$\to$9 spectra in the
  inset (histograms) are shown at spectral resolutions of 9.69, 19.55,
  19.55, 25.00, 15.63, and 15.63\,MHz (84, 68, 56, 54, 30, and
  27\,\kms ), respectively. The black curves in each panel show
  Gaussian fits to the line emission or absorption. \label{f1}}
%
\end{figure*}

\begin{figure*}[tbh]
\epsscale{1.15}
\plotone{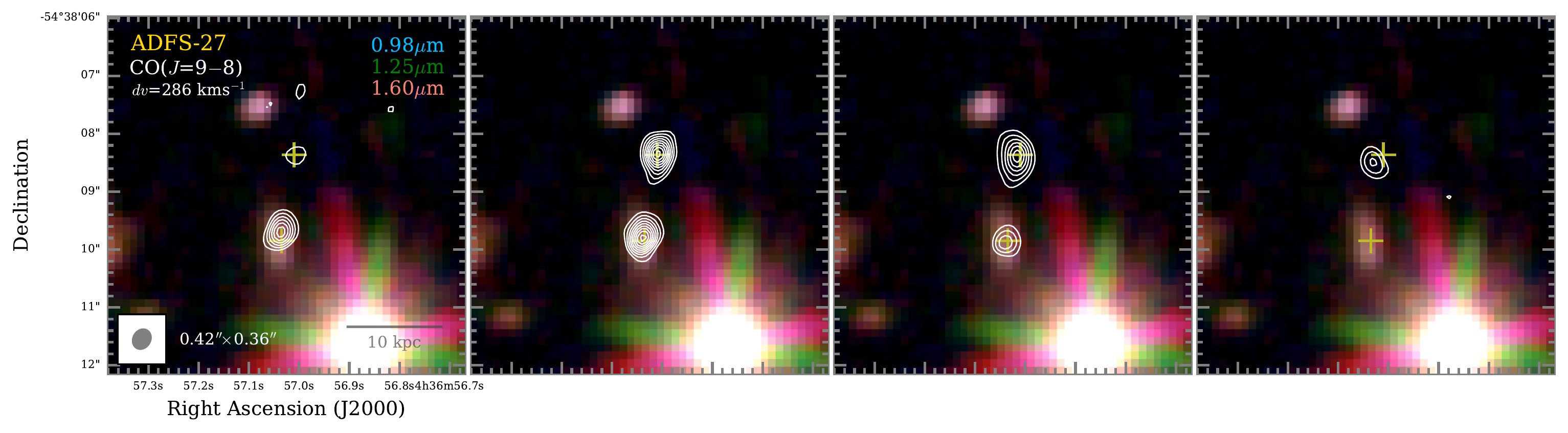}
\plotone{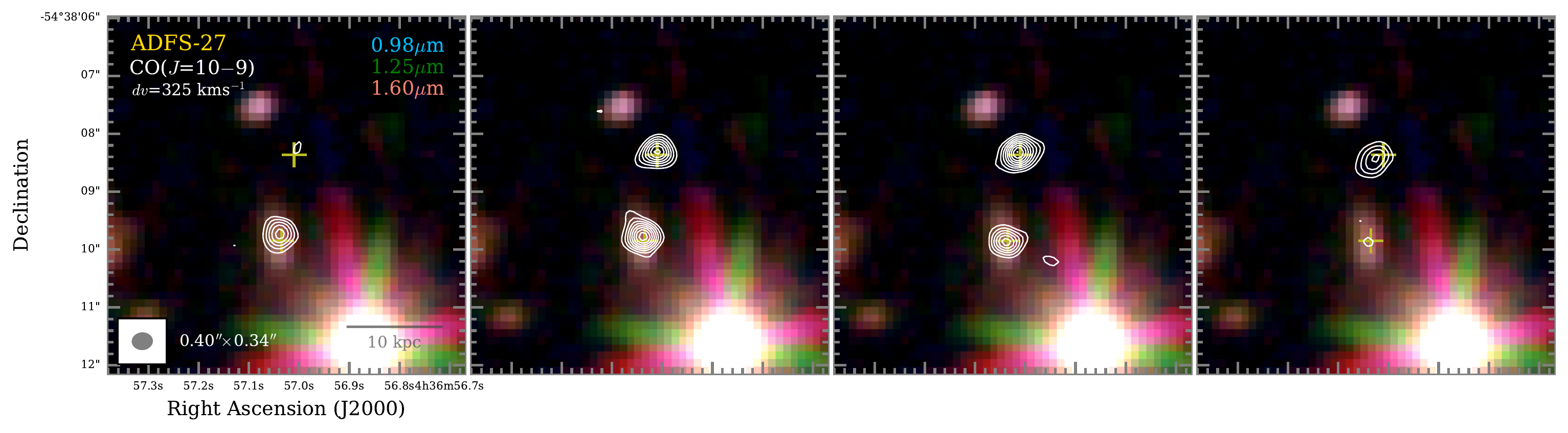}

\vspace{-2mm}

\caption{Velocity channel maps of the continuum-subtracted CO
  $J$=9$\to$8 (top) and 10$\to$9 emission (bottom), imaged with robust
  0.5 weighting over 286 and 325\,\kms\ wide bins (starting at
  --503 and --707\,\kms ), and overlaid on the same images as in
  Fig.~\ref{f3}. Beam sizes are indicated in the bottom left corners
  of the bluemost velocity channel maps. The crosses indicate the peak
  positions of the 0.87\,mm continuum emission. Contours are shown in
  steps of 2$\sigma$ (where 1$\sigma$=46 and
  61\,$\mu$Jy\,beam$^{-1}$), starting at $\pm$4$\sigma$.
  \label{f4}}
%
\end{figure*}

\subsection{Line Emission and Absorption}

We here report our new findings on CO, H$_2$O, OH$^+$, and CH$^+$
emission and absorption toward \adfs27. The main results are
summarized in Table~\ref{t2}.

\subsubsection{CO Emission}

\begin{figure*}
\epsscale{0.44}
\plotone{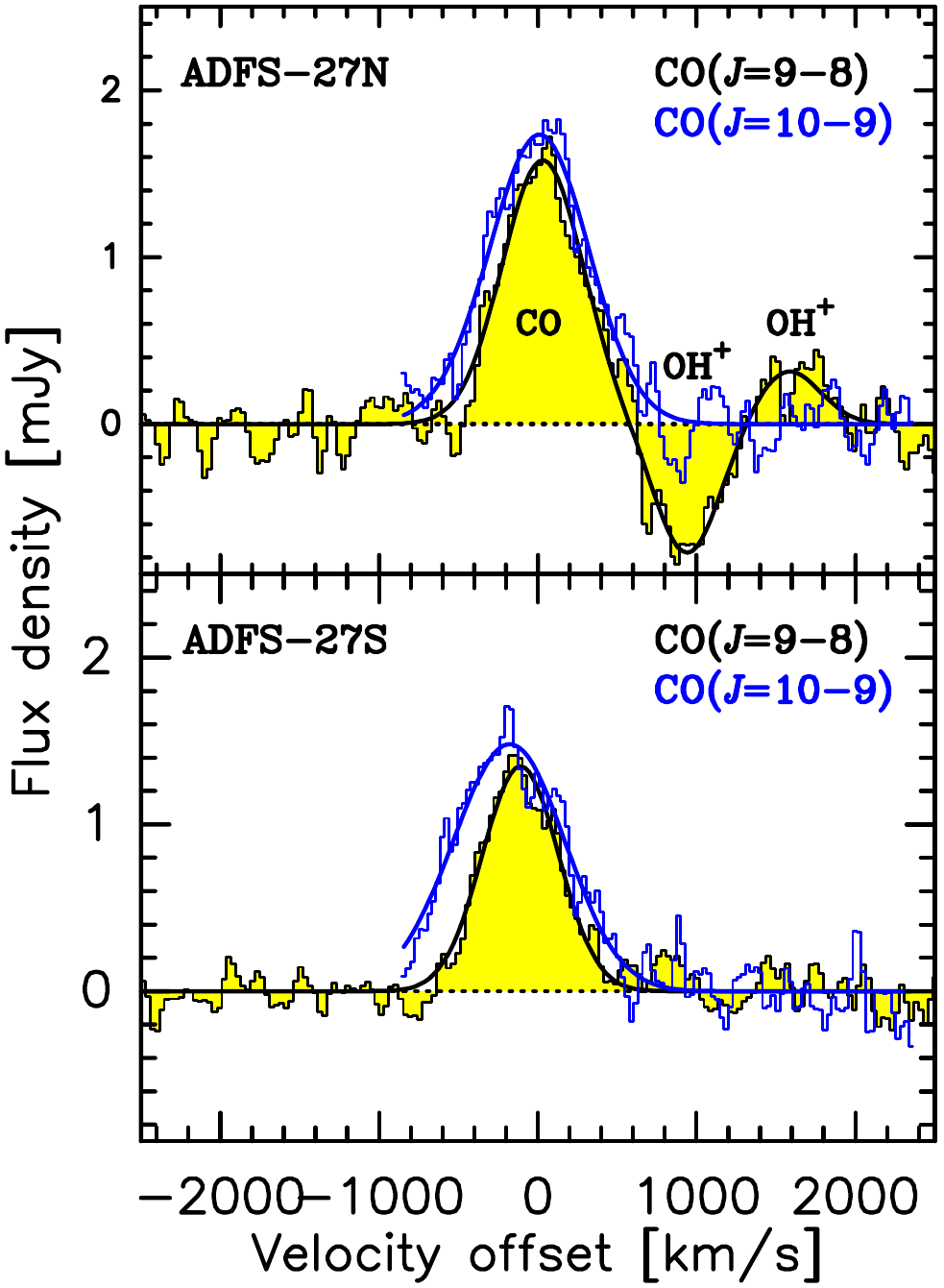}
\plotone{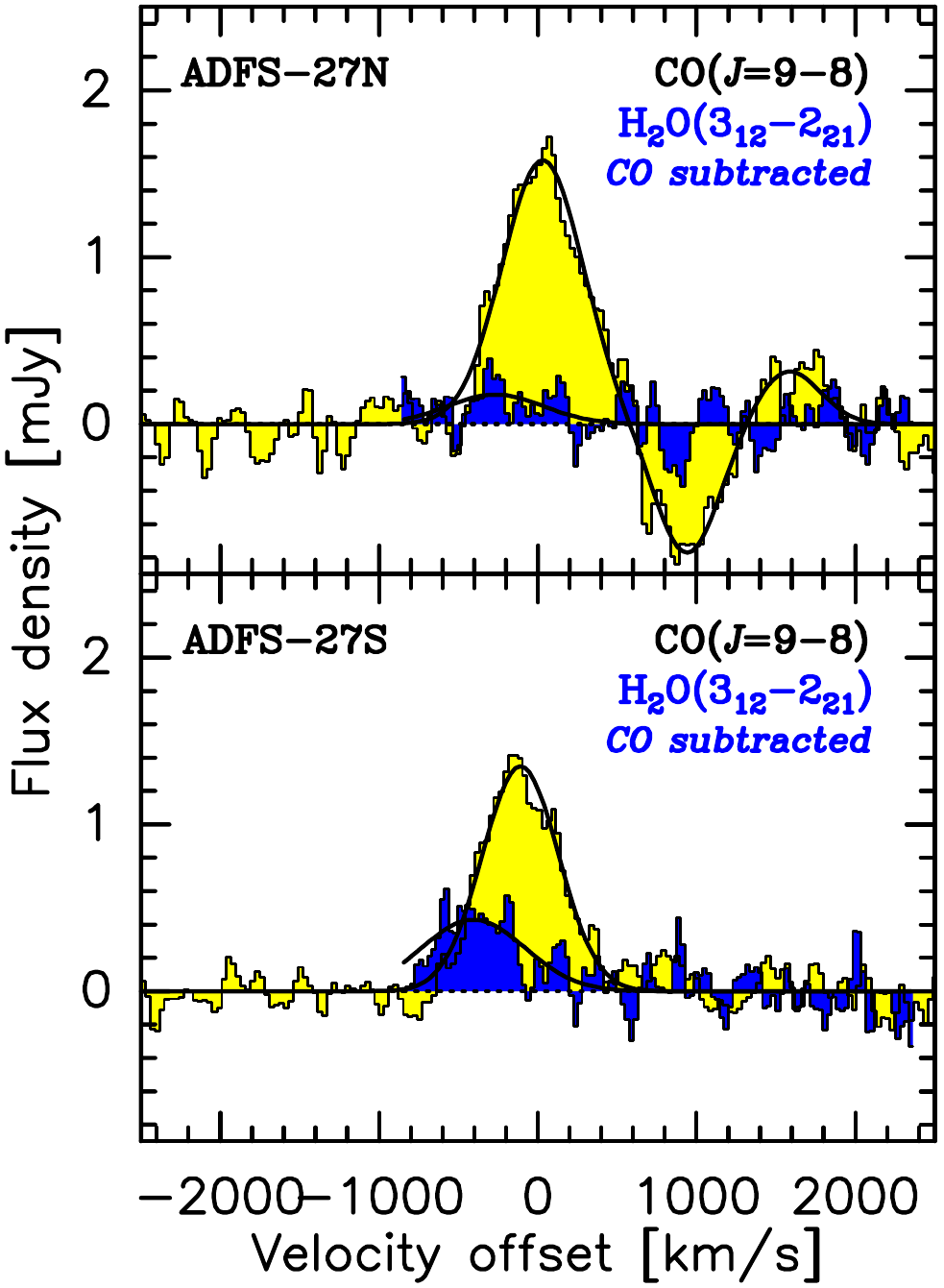}

\vspace{-2mm}

\caption{Comparison of the CO $J$=9$\to$8 (black/yellow histograms)
  and 10$\to$9 (blue) line profiles toward ADFS-27N (top left) and S
  (bottom left), and spectra of the \waterb\ line profiles (blue)
  after subtracting \jco\ emission (right). The spectra are shown at
  the same resolution as in Fig~\ref{f1}, and continuum emission has
  been subtracted in all cases. Left:\ By comparison, the red line
  wing of the \ico\ line does not appear to be significantly affected
  by the OH$^+$ absorption component in ADFS-27N. On the other hand,
  the \jco\ line in ADFS-27S appears to be significantly more
  broadened compared to \ico, due to a stronger contribution from the
  H$_2$O 3$_{12}$$\to$2$_{21}$ line, which is supported by the fits (black
  and blue lines) to the line profiles when assuming a common redshift
  for CO $J$=9$\to$8 and 10$\to$9. Right:\ The H$_2$O emission is
  weaker than \jco\ in both source components, but the relative
  strength is higher in the less dust-obscured ADFS-27S.
 \label{f5a}}
%
\end{figure*}

We detect CO $J$=2$\to$1 and 8$\to$7 emission toward \adfs27\ at
7.6$\sigma$ and 18$\sigma$ peak significance (Fig.~\ref{f2}),
respectively. We also detect and spatially resolve CO $J$=9$\to$8 and
10$\to$9 emission towards ADFS-27N and S at 23$\sigma$ peak
significance in each case (naturally-weighted maps are shown in
Fig.~\ref{f3}; CO $J$=10$\to$9 is not corrected for a subdominant
contribution from the H$_2$O 3$_{12}$$\to$2$_{21}$ line). The line
FWHM for the CO $J$=2$\to$1 and 8$\to$7 lines is (910$\pm$143) and
(743$\pm$68)\,\kms, respectively, as compared to (651$\pm$59) and
(710$\pm$103)\,\kms\ for the previously-detected CO $J$=5$\to$4 and
6$\to$5 lines. The CO $J$=9$\to$8 and 10$\to$9 lines summed over both
source components yield FWHM values of (597$\pm$18) and
(816$\pm$16)\,\kms, where the latter is not corrected for any
contribution from the H$_2$O 3$_{12}$$\to$2$_{21}$ line. The overall
trend may suggest changes in the line ratio between ADFS-27N and S
from the CO $J$=5$\to$4 to the 8$\to$7 transition due to a potential
increase in overall line widths, but the values are consistent within
the uncertainties. On the other hand, the CO $J$=9$\to$8 line may be
somewhat narrower than expected, possibly due to a contribution from
the nearby OH$^+$ absorption feature, while the CO $J$=10$\to$9 line
appears broader than all lines except CO $J$=2$\to$1 due to a likely
contribution from the H$_2$O 3$_{12}$$\to$2$_{21}$ line. To further
investigate these issues, we compare the profiles of the CO
$J$=9$\to$8 and 10$\to$9 lines for each component individually in
Fig.~\ref{f5a}. The consistent red wing between both lines does not
suggest significant reduction of the \ico\ line width or flux due to
OH$^+$ absorption in either source component. On the other hand, the
extended blue wing of the \jco\ line in both components suggests a
significant contribution from the H$_2$O 3$_{12}$$\to$2$_{21}$ line in
both cases. To de-blend the CO and H$_2$O lines, we fix the line
centroids of both lines to the high signal-to-noise ratio \ico\ lines,
and we assume a common width of the CO and H$_2$O lines. The
CO-subtracted H$_2$O spectra are shown in the right panels of
Fig.~\ref{f5a} for reference, and the H$_2$O line parameters are
studied below. We also simultaneously fit the \ico\ and OH$^+$
emission, and report the OH$^+$ line parameters below. For the CO
$J$=9$\to$8 and 10$\to$9 lines, we find line FWHM values of
(599$\pm$28) and (711$\pm$22)\,\kms\ for ADFS-27N, and (565$\pm$18)
and (767$\pm$28)\,\kms\ for ADFS-27S, respectively. ADFS-27N is
redshifted by (140$\pm$13)\,\kms\ relative to ADFS-27S in the
\ico\ line, which suggests that the lines appear broadened in
spatially-unresolved measurements due to the internal kinematic
structure of the merging system. We find redshifts of
$z$=5.65568$\pm$0.00023 and 5.65258$\pm$0.00016 for ADFS-27N and S,
respectively. Intensities and luminosities for all lines are reported
in Table~\ref{t2}.

\begin{figure*} 
\begin{deluxetable}{ l c l c c c }

\tablecaption{Line fluxes and luminosities 
in \adfs27. \label{t2}}
\tablehead{
Transition & $I_{\rm line}$ & dv$_{\rm FWHM}$ & $L'_{\rm line}$ & $L_{\rm line}$ & $r_{J2}$\tablenotemark{a} \\
           & (Jy\,\kms ) & (\kms ) & (10$^{10}$\,K\,\kms\,pc$^2$) & (10$^8$\,\lsol ) }
\startdata
\bco\   & 0.716 $\pm$ 0.087 & 910$\pm$143 & 19.93 $\pm$ 1.42 & 0.782 $\pm$ 0.095 & 1 \\
\eco\   & 2.68 $\pm$ 0.20 & 651$\pm$59 & 11.96 $\pm$ 0.92 & 7.32 $\pm$ 0.56 & 0.60 $\pm$ 0.09 \\ 
\fco\   & 2.82 $\pm$ 0.34 & 710$\pm$103 & 8.73 $\pm$ 1.07 & 9.24 $\pm$ 1.13 & 0.44 $\pm$ 0.07 \\ 
\hco\   & 2.24 $\pm$ 0.11 & 743$\pm$68 & 3.90 $\pm$ 0.19 & 9.78 $\pm$ 0.48 & 0.20 $\pm$ 0.03 \\
\ico\   & 1.80 $\pm$ 0.04 & 599$\pm$28 (N) & 2.48 $\pm$ 0.06 & 8.84 $\pm$ 0.21 & 0.124 $\pm$ 0.015\\
        &                 & 565$\pm$18 (S)    &                 &                 & \\
\jco\   & 2.18 $\pm$ 0.05 & 711$\pm$22 (N) & 2.43 $\pm$ 0.06 & 11.89 $\pm$ 0.29 & 0.121 $\pm$ 0.015 \\
        &                 & 767$\pm$28 (S)    &                 &                 & \\
CH$^+$($J$=1$\to$0) abs.\tablenotemark{b} & {\em --0.36$\pm$0.14} & \,\,\,\,\,\,\,\,\,\,\,\,\,\,\,\,\,\,\,\,\,(N) & & & \\
OH$^+$(1$_1$$\to$0$_1$) abs.\ & --0.296 $\pm$ 0.027 & 422$\pm$50 (N) & & & \\
OH$^+$(1$_1$$\to$0$_1$) em.\ & 0.147$\pm$0.018 & 745$\pm$87 (N) & 0.204 $\pm$ 0.025 & 0.719 $\pm$ 0.088 & \\
OH$^+$(1$_2$$\to$0$_1$) em.\ & 0.17$\pm$0.04 &$\sim$740 (N) & 0.27 $\pm$ 0.06 & 0.79 $\pm$ 0.17 & \\
\water\tablenotemark{b} & {\em 0.83 $\pm$ 0.22} & 503$\pm$163 & {\em 2.17 $\pm$ 0.58} & {\em 2.96 $\pm$ 0.80} & \\
\waterb\tablenotemark{c} & 0.48 $\pm$ 0.06 & ({\em fixed}; N/S) & 0.54 $\pm$ 0.07 & 2.63 $\pm$ 0.33 & \\
\waterc\tablenotemark{b} & {\em 1.26 $\pm$ 0.32} & 1150$\pm$340 (S) & {\em 1.38 $\pm$ 0.35} & {\em 6.96 $\pm$ 1.75} & \\
\enddata
\tablenotetext{\rm a}{Line brightness temperature ratio relative to \bco. For reference, the large velocity gradient (LVG) modeling suggests $r_{21}$=0.95 between CO $J$=2$\to$1 and 1$\to$0.}
\tablenotetext{\rm b}{Tentative detection, independent confirmation required.}
\tablenotetext{\rm c}{De-blended from \jco\ line assuming a common line width.}
\end{deluxetable}
\end{figure*}


\subsubsection{H$_2$O Emission}

We tentatively detect H$_2$O(3$_{21}$$\to$3$_{12}$) emission at
approximately 4$\sigma$ significance, at a position consistent with
ADFS-27S. The formal FWHM of the line is (1150$\pm$340)\,\kms, which
is consistent with the width of the CO lines within the
uncertainties. Through de-blending from the CO($J$=10$\to$9) line, we
also detect H$_2$O(3$_{12}$$\to$2$_{21}$) emission toward ADFS-27S,
and at lower significance, toward ADFS-27N. The combined detection
significance of both components is about 8$\sigma$ when neglecting
systematic uncertainties in the de-blending process. The line widths
are tied to the \jco\ line, and a simultaneous fit tied to the
\ico\ redshift of each component gives FWHM of (711$\pm$22) and
(767$\pm$28)\,\kms\ for ADFS-27N and S, respectively. Given the H$_2$O
de-excitation cascade process expected for radiatively-excited H$_2$O
lines (as is likely the case for ADFS-27), the
H$_2$O(3$_{21}$$\to$3$_{12}$) line is expected to be brighter than the
H$_2$O(3$_{12}$$\to$2$_{21}$) line based on photon number conservation
arguments (see, e.g., Gonzalez-Alfonso et al.\ \citeyear{ga10}). We
find an H$_2$O 3$_{12}$$\to$2$_{21}$/3$_{21}$$\to$3$_{12}$ ratio of
$r_{\rm w}$=0.39$\pm$0.11 (Table~\ref{t2}), which is formally
consistent with a 1:2 ratio within the uncertainties. As such, we
would expect the H$_2$O(3$_{12}$$\to$3$_{03}$) line, as the second
branch of the cascade, to be detectable at a strength comparable to
the H$_2$O(3$_{12}$$\to$2$_{21}$) line in \adfs27. Additionally, a
detection of H$_2$O(3$_{12}$$\to$2$_{21}$) and
H$_2$O(3$_{21}$$\to$3$_{12}$) is consistent with the previous
tentative detection of H$_2$O(2$_{11}$$\to$2$_{02}$) in this source
(Riechers et al.\ \citeyear{riechers17}). While tentative, the peak
velocity of that line also appears to be blueshifted compared to the
CO emission centroid of the entire source, which is consistent with
stronger H$_2$O emission in ADFS-27S (as the blueshifted source
component) than in ADFS-27N. The comparatively brighter H$_2$O
emission in ADFS-27S is consistent with its warmer dust spectral
energy distribution shape, under the assumption that the H$_2$O line
ladder is dominantly excited radiatively. This is because warmer dusts
leads to a greater availability of 75.4\,$\mu$m photons, which pump
the ortho-H$_2$O 3$_{21}$ level from the 2$_{12}$ state (see, e.g.,
Gonzalez-Alfonso et al.\ \citeyear{ga12}; Riechers et
al.\ \citeyear{riechers13b}). Higher signal-to-noise ratio detections
of spectrally isolated H$_2$O lines are required for a more detailed
interpretation.

\begin{figure*}
\epsscale{0.8}
\plotone{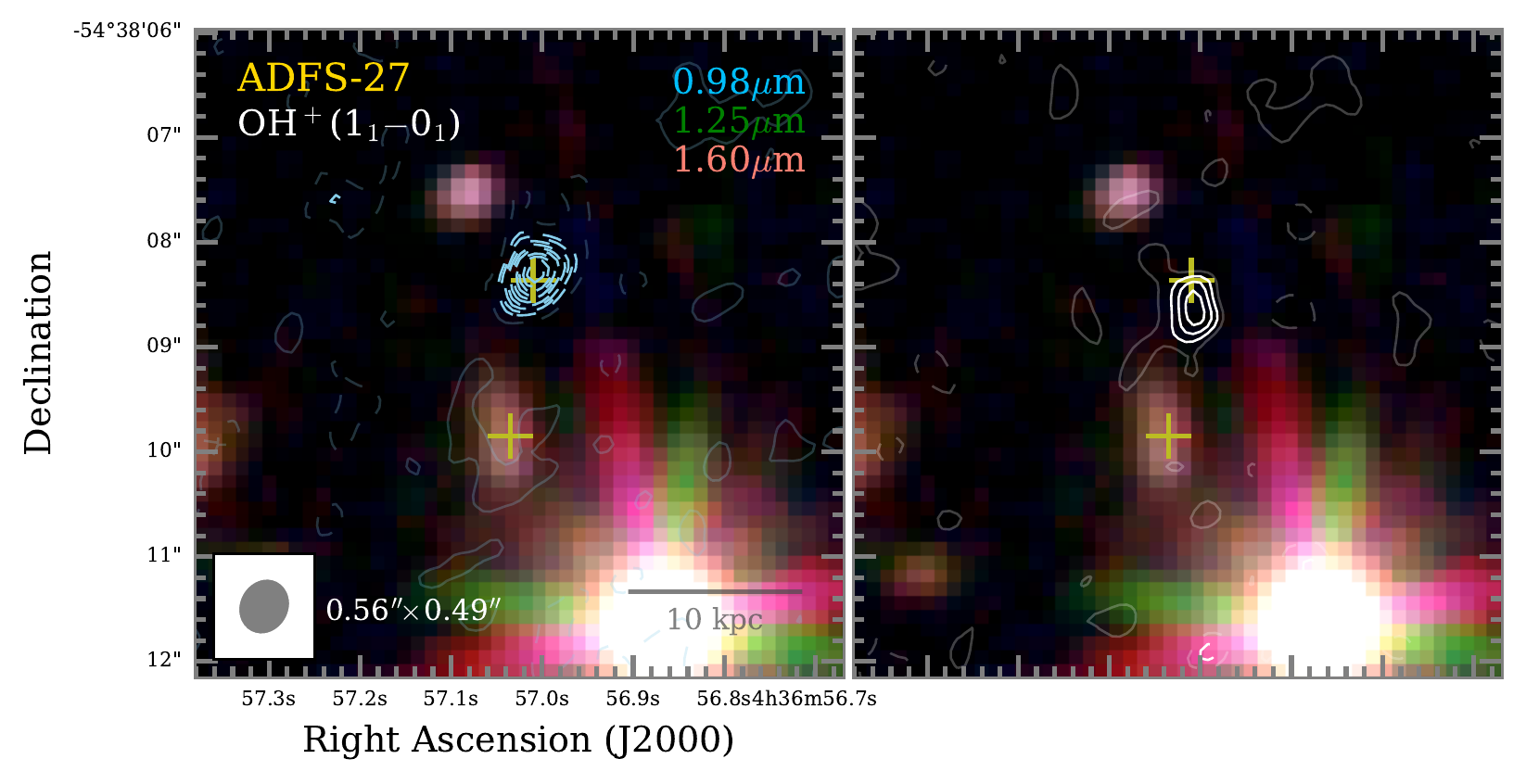}
\epsscale{0.31}
\plotone{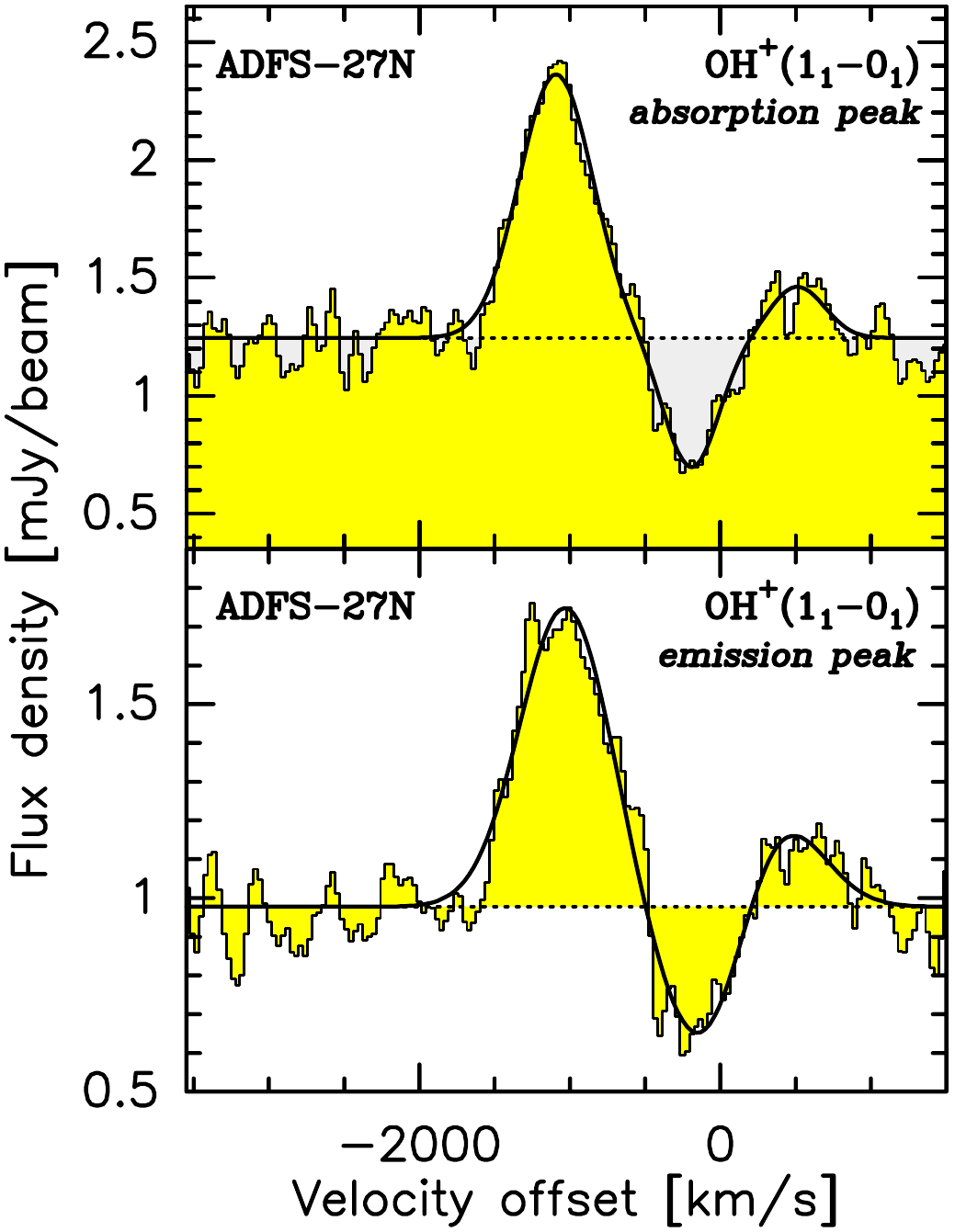}

\vspace{-2mm}

\caption{Maps of the continuum-subtracted OH$^+$(1$_1$$\to$0$_1$)
  absorption (left) and emission (middle) toward \adfs27, overlaid on
  the same images as in Fig.~\ref{f3}, and spectra including continuum
  emission at the peak positions (right). The beam size is shown in
  the bottom left corner of the left panel. Left/middle:\ The crosses
  indicate the same positions as in Fig.~\ref{f4}. The absorption
  (emission) components are shown at spectral resolutions of 664 and
  815\,\kms\ (343.75 and 421.875\,MHz), respectively. Contours are
  shown in steps of $\pm$1$\sigma$=0.019 and
  0.023\,Jy\,\kms\,beam$^{-1}$, respectively, starting at
  $\pm$2$\sigma$. Right:\ Spectra (histograms) were extracted at the
  peak positions in the maps, and are shown at a resolution of
  30\,\kms\ (15.63\,MHz). The black curves show Gaussian fits to the
  line emission and absorption, including the \ico\ line. \label{f6}}
%
\end{figure*}

\begin{figure*}
\epsscale{0.6}
\plotone{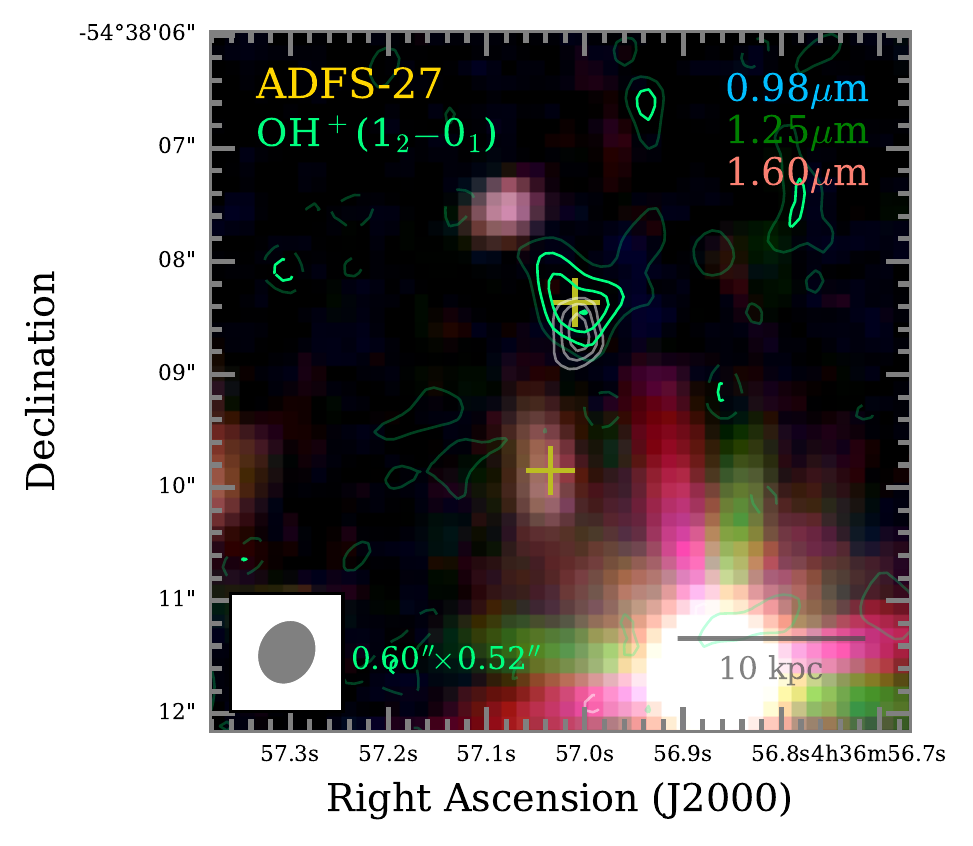}
\epsscale{0.4}
\plotone{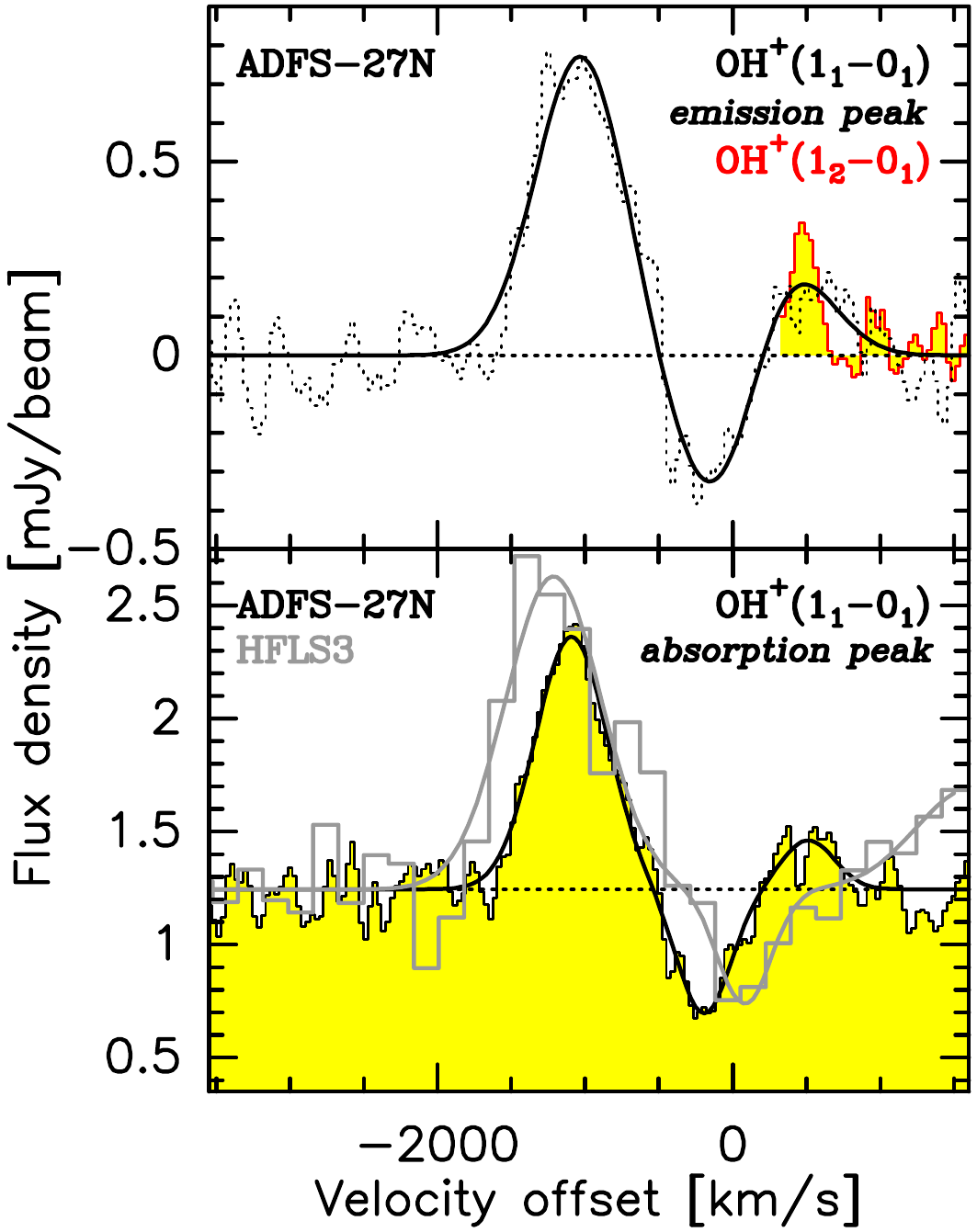}

\vspace{-2mm}

\caption{Maps of the continuum-subtracted OH$^+$(1$_2$$\to$0$_1$)
  emission (left) toward \adfs27, overlaid on the same images as in
  Fig.~\ref{f3}, spectrum at the peak position (top right), and
  comparison of the OH$^+$(1$_1$$\to$0$_1$) profiles in \adfs27\ and
  HFLS3 (bottom right). The beam size is shown in the bottom left
  corner of the left panel. Left:\ The crosses indicate the same
  positions as in Fig.~\ref{f4}. The OH$^+$(1$_2$$\to$0$_1$) emission
  (green contours) is shown over a spectral bandwidth of
  321\,\kms\ (156.25\,MHz). Contours are shown in steps of
  1$\sigma$=0.0135\,Jy\,\kms\,beam$^{-1}$, starting at
  $\pm$2$\sigma$. The OH$^+$(1$_1$$\to$0$_1$) emission (white) is
  shown for comparison, using the same contour levels as in
  Fig.~\ref{f6}, but omitting the $\pm$2$\sigma$ levels for clarity.
  Top right:\ OH$^+$(1$_2$$\to$0$_1$) spectrum (red histogram)
  extracted at the peak position in the map is shown at a resolution
  of 32\,\kms\ (15.63\,MHz). The same OH$^+$(1$_1$$\to$0$_1$) spectrum
  (dotted histogram) and fit (curve) as in Fig.~\ref{f6}, bottom
  right, are shown for comparison, except for a subtraction of the
  continuum emission. The difference in spectral line shapes is not
  statistically significant.  Bottom right: Same as Fig.~\ref{f6}, top
  right, but with the same spectral features in the $z$=6.34 DSFG
  HFLS3 overlaid (gray histogram; Riechers et
  al.\ \citeyear{riechers13b}). The HFLS3 data have been normalized to
  the same continuum flux for clarity, and are shown at a spectral
  resolution of 170\,\kms\ (80\,MHz). The OH$^+$ feature appears more
  redshifted in HFLS3, and the absorption component peaks closer
  to the systemic velocity, but it may affect the red wing of the
  \ico\ emission line. The peak strengths of the absorption are
  similar in both cases.\label{f6b}}
%
\end{figure*}

\subsubsection{OH$^+$ Emission/Absorption}

We detect a P-Cygni shaped absorption/emission profile of the
OH$^+$(1$_1$$\to$0$_1$) line toward ADFS-27N, while no OH$^+$ line is
detected toward ADFS-27S (Fig.~\ref{f6}). The OH$^+$ absorption and
emission components in ADFS-27 are detected at peak significances of
14$\sigma$ and 5.6$\sigma$, respectively. The OH$^+$ absorption
component is blueshifted by (216$\pm$16)\,\kms\ at its map peak
position relative to the \ico\ line, while the emission component is
redshifted by (171$\pm$10)\,\kms\ at its peak position relative to the
\ico\ line (or by (233$\pm$10)\,\kms\ relative to the CO $J$=9$\to$8
emission at the CO and continuum peak position). The OH$^+$ absorption
component is spatially coincident with the continuum emission within
0.076$''$ ($\lesssim$0.5\,kpc), while the peak of the emission
component is offset to the south by 0.29$''$ (1.7\,kpc). The OH$^+$
absorption and emission components have FWHM values of (422$\pm$50)
and (745$\pm$87)\,\kms, respectively. Thus, the emission component has
a width similar to the high-$J$ CO lines, but the absorption component
is narrower.

A comparison to the OH$^+$(1$_1$$\to$0$_1$) line detected toward HFLS3
($z$=6.34; Riechers et al.\ \citeyear{riechers13b}) shows a comparable
absorption strength, but the absorption and emission components in
HFLS3 appear significantly more redshifted (Fig.~\ref{f6b}; only the
red wing of the emission component is seen in HFLS3 due to the
limited coverage of the bandpass).

We also detect the emission wing of the OH$^+$(1$_2$$\to$0$_1$)
emission feature (Fig.~\ref{f1} and \ref{f6b}) at 5$\sigma$ peak
significance. The highest signal-to-noise ratio is obtained when
averaging the emission over 320\,\kms, but the line remains
detected at the $>$4$\sigma$ level when averaging over the full range
showing positive emission in the spectrum (740\,\kms ). The peak
of the emission also shifts to the south by a fraction of the beam
size when averaging over the broader velocity range, but remains
consistent with the position shown in Fig.~\ref{f6b} within the
uncertainties. Overall, we find that the peak positions, redshifts,
and line profiles of the OH$^+$ 1$_2$$\to$0$_1$ and 1$_1$$\to$0$_1$
emission components agree within the uncertainties. Due to the fact
that only part of the line profile is seen, we only report an
integrated flux for the OH$^+$ 1$_2$$\to$0$_1$ emission component, and
we focus our main analysis on the OH$^+$(1$_1$$\to$0$_1$) line.

\begin{figure*}
\epsscale{0.6}
\plotone{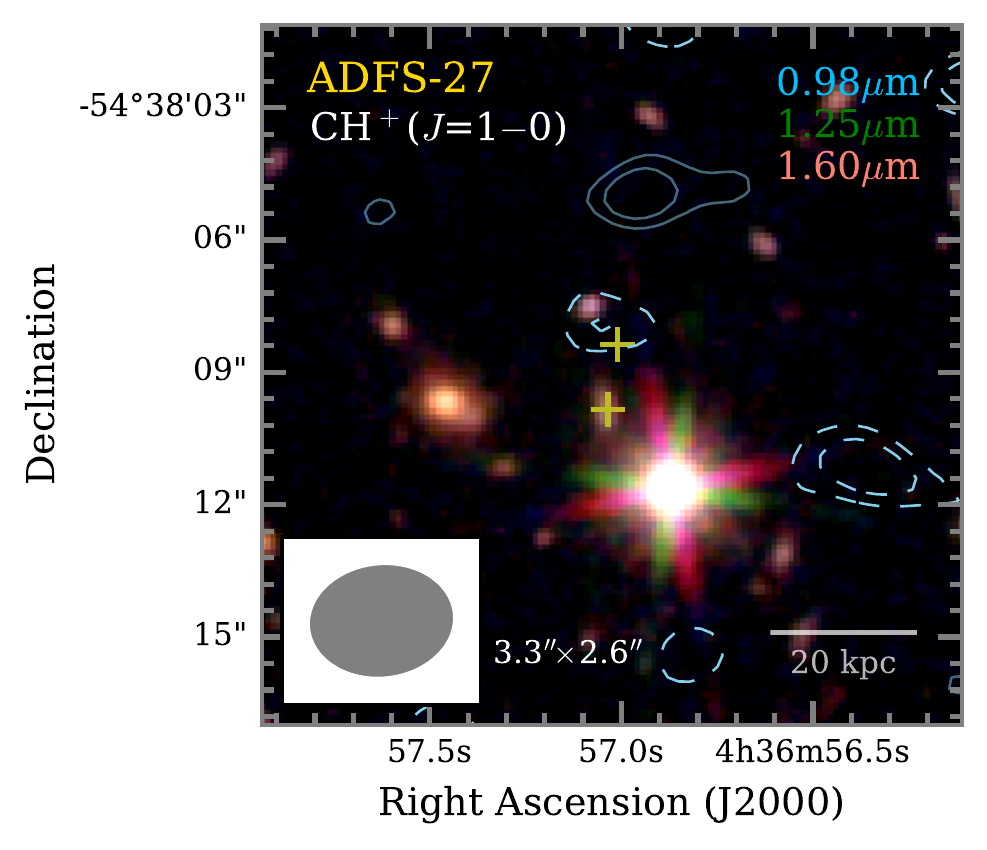}
\epsscale{0.4}
\plotone{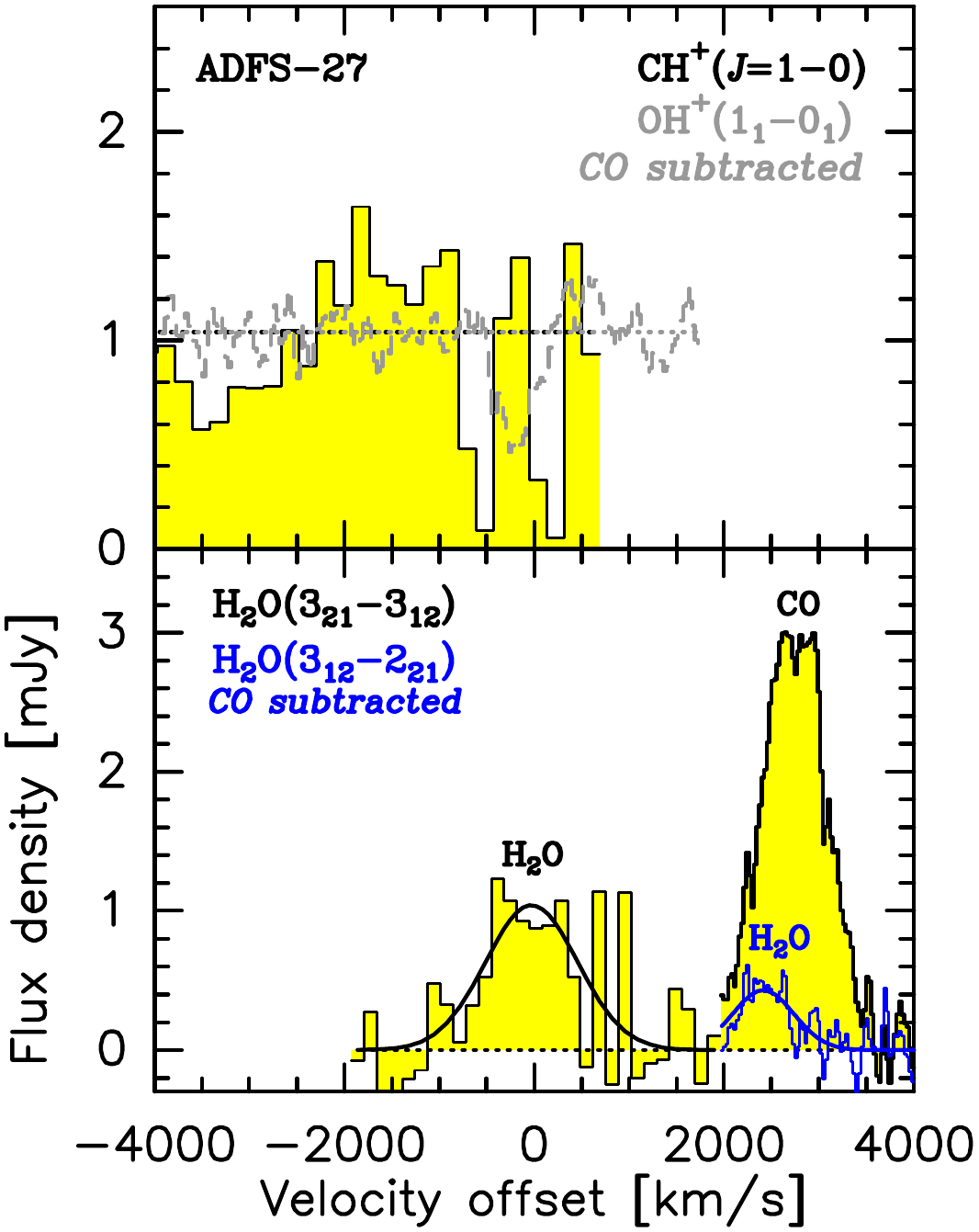}

\vspace{-2mm}

\caption{Moment-0 map (left) over 1100\,\kms\ (460.938\,MHz) and
  spectrum of CH$^+$ $J$=1$\to$0 (yellow/black histogram; top right),
  and H$_2$O 3$_{21}$$\to$3$_{12}$ (bottom right) toward \adfs27. The
  CH$^+$ and H$_2$O spectra are shown at 78\,MHz (187 and 134\,\kms )
  resolution, respectively. Left:\ Contours are shown at the
  $\pm$2.0$\sigma$ and $\pm$2.5$\sigma$ level, where
  1$\sigma$=0.143\,Jy\,\kms. The tentative CH$^+$ absorption feature
  appears to be associated with ADFS-27N, which is also the source
  component showing OH$^+$ absorption. Top right:\ The
  OH$^+$(1$_1$$\to$0$_1$) line profile of ADFS-27N is shown for
  comparison (dashed histogram), after subtraction of the
  \ico\ emission. The CH$^+$ spectrum (solid histogram) is consistent
  with absorption, but the current signal-to-noise ratio is not
  sufficient to claim a solid detection. Bottom right:\ Since the
  peak position of the H$_2$O 3$_{21}$$\to$3$_{12}$ line is consistent with
  ADFS-27S, its \jco-subtracted H$_2$O 3$_{12}$$\to$2$_{21}$
  spectrum (blue) is shown for comparison.
 \label{f5b}}
%
\end{figure*}

\subsubsection{CH$^+$ Absorption}

We tentatively detect the CH$^+$($J$=1$\to$0) line (Fig.~\ref{f5b}) at
2.5$\sigma$ significance in absorption at a position consistent with
ADFS-27N. The feature is formally consistent with the central velocity
and strength of the OH$^+$(1$_1$$\to$0$_1$) absorption detected toward
the same source component (see also Table~\ref{t2}), but a broader
feature cannot be ruled out with the current data. Given the limited
signal-to-noise ratio of the detection, we do not report parameters
from a fit to the line profile. More sensitive observations are
required to confirm this feature, and to study its properties in more
detail.

\section{Analysis and Discussion}

We here determine a broad range of physical properties for \adfs27,
to place them in the more general context required to improve our
understanding of galaxy evolution. The main quantities are summarized
in Tables \ref{t3}, \ref{t4}, and \ref{t5}.

\subsection{Spectral Energy Distribution Modeling}

\subsubsection{{\sc CIGALE}}

We have used the Code Investigating GALaxy Emission ({\sc CIGALE};
Burgarella et al.\ \citeyear{Burgarella2005}; Noll et
al.\ \citeyear{Noll2009}; Boquien et al.\ \citeyear{Boquien2019})
spectral energy distribution (SED) fitting package to model the full
SED of \adfs27. We have run the code on all spatially-integrated
photometry from Table~\ref{t1} for the entire galaxy, and also on the
two source components ADFS-27N and S individually, using only the
photometry in those bands where the emission is resolved into both
sources. We ran two series of fits, with star-formation histories
(SFHs) either limited to approximately the age of the universe at
$z$=5.655 (i.e., $\leq$1.0\,Gyr), or to an age of $\leq$0.2\,Gyr, as
is characteristic for young dusty starbursts (e.g., Greve et
al.\ \citeyear{greve05}; Bergvall et al.\ \citeyear{bergvall16}). We
sampled standard ranges for all main parameters within {\sc CIGALE},
and used a Bruzual \& Charlot (\citeyear{bc03}) single stellar
population and a Chabrier (\citeyear{chabrier03}) stellar initial mass
function with a ``delayed'' SFH (i.e., where star-formation rate
SFR($t$) $\propto$ $t$/$\tau^2$ $\times$ exp($-t$/$\tau$), and the fit
parameter $\tau$ is the time at which the SFR peaks; Boquien et
al.\ \citeyear{Boquien2019}; see also Burgarella et
al.\ \citeyear{burgarella20}) for all fits. We fixed the power-law
slope of the dust attenuation law to --0.7 for both the interstellar
medium and birth clouds, and assumed a polycyclic aromatic hydrocarbon
(PAH) mass fraction (which is not directly constrained by the data) of
$q_{\rm PAH}$=3.9\% (i.e., similar to dust found in the Milky Way and
nearby galaxies with near-solar metallicity; e.g., Draine \& Li
\citeyear{dl07}, and references therein).\footnote{We allowed for a
  range in metallicity in our fits, but the metallicity remains
  difficult to constrain directly without rest-frame optical
  spectroscopy.} This is consistent with the range of $q_{\rm
  PAH}$=0.47\%--3.9\% found for a sample of infrared-luminous galaxies
at $z$=0.5--4 (Magdis et al.\ \citeyear{magdis12}).\footnote{The
  uncertainty due to this assumption is fully captured by the error
  bars for the most relevant parameters, like $M_{\rm dust}$.}  For
ADFS-27 as a whole and for ADFS-27S (which is detected in the
near-infrared bands), the best-fit parameters for both series agree
within the uncertainties. We thus adopt those from the series with
less constraints on the SFH in the following, and only use those from
the other series in the evaluation of the true uncertainties. For
ADFS-27N (which is not detected in the near-infrared bands), the
series with more stringent constraints on the SFH appears to provide
more reasonable results, and thus, are adopted in the following.

We find a total dust luminosity of $L_{\rm
  dust}$=(2.62$\pm$0.21)$\times$10$^{13}$\,\lsol\ and a SFR of
(3100$\pm$330)\,\msol\,yr$^{-1}$ for \adfs27. We also find a total
dust mass of $M_{\rm
  dust}$=(4.19$\pm$0.39)$\times$10$^9$\,\msol. These values would
differ by only 0.6\%, 3\%, and --1\% when adopting the more restricted
parameter study, which is indistinguishable within the
uncertainties. The dust mass is consistent with that found by Riechers
et al.\ (\citeyear{riechers17}) within the uncertainties.

{\sc CIGALE} uses the simplified Draine \& Li (\citeyear{dl07})
prescription for which dust is heated by starlight in an intensity
range with $U_{\rm min}$$<$$U$$<$$U_{\rm max}$, where $U$ is a
dimensionless scaling factor of the interstellar radiation field in
the solar neighborhood as estimated by Mathis et
al.\ (\citeyear{mathis83}), and $U_{\rm max}$=10$^6$ is fixed to the
value found for nearby star-forming galaxies. We find $U_{\rm
  min}$=32$\pm$11, and a power-law slope of $\alpha$=2.8$\pm$0.3 for
the distribution of starlight intensities $dM_{\rm
  dust}$/$dU$$\propto$$U^{-\alpha}$ (e.g., Dale et
al.\ \citeyear{dale01}). A fraction (1--$\gamma$)$\simeq$60\%$\pm$30\%
of the dust is heated by starlight at $U_{\rm min}$ (representing the
diffuse interstellar medium component), and
$\gamma$$\simeq$40\%$\pm$30\% is heated by more intense starlight that
follows the power-law distribution. Following Draine \& Li
(\citeyear{dl07}), this corresponds to $f_{\rm PDR}$$\simeq$77\%
of the dust heating\footnote{Dust heating is defined here as the
  fraction of $L_{\rm dust}$ radiated by dust grains in regions where
  $U$$>$100, i.e., $\sim$3$\times$$U_{\rm min}$.} being due to
reprocessed starlight from photon-dominated regions (PDRs), rather
than the interstellar radiation field. This would be consistent with a
picture in which young starbursts are hosted by a largely fractured,
clumpy interstellar medium, embedded in the diffuse interstellar
medium that fills most of the volume of the sources. We note however
that $\gamma$ is not constrained at high significance by the current
data, and thus is the main source of uncertainty in our interpretation.

We find that, despite the large quantities of dust present, the
stellar light detected toward ADFS-27S appears to suffer only a modest
extinction, corresponding to an extinction
$A_V$$\simeq$(1.0$\pm$0.1)\,mag, while ADFS-27N appears to suffer a
substantially larger extinction $A_V$$\simeq$(5.1$\pm$0.8)\,mag. We
find young ages of (110$\pm$60) and (80$\pm$30)\,Myr for the stellar
populations of ADFS-27N and S, respectively (which are consistent with
the picture described in the previous paragraph), and total stellar
masses of (1.6$\pm$1.3) and
(0.46$\pm$0.15)$\times$10$^{11}$\,\msol. Based on our findings from
both fit series, we estimate that the stellar mass $M_\star$ for
ADFS-27N is uncertain by at least a factor of three in practice. The
individual dust masses are $M_{\rm dust}$=(2.60$\pm$0.38) and
(1.46$\pm$0.20)$\times$10$^9$\,\msol, respectively. This suggests that
the dust yield from the stellar populations is of order 2\% after
about 100\,Myr, i.e., after sufficient time for stars of
$>$5--8\,\msol\ produced early on to end their life cycles and enrich
their surroundings (assuming that no additional, older stellar
populations that have produced the bulk of the dust are present).

\begin{figure*}
\epsscale{0.58}
\plotone{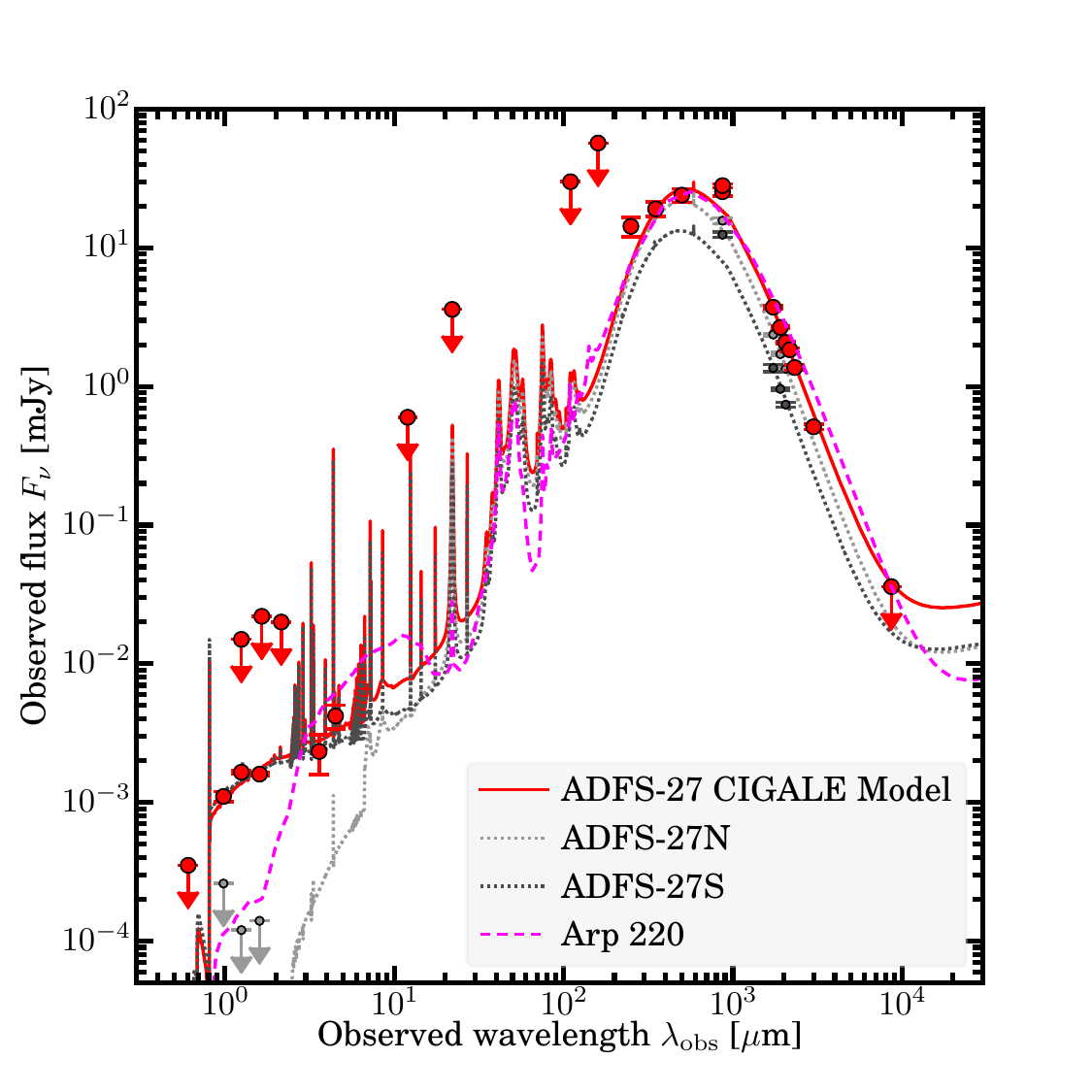}
\plotone{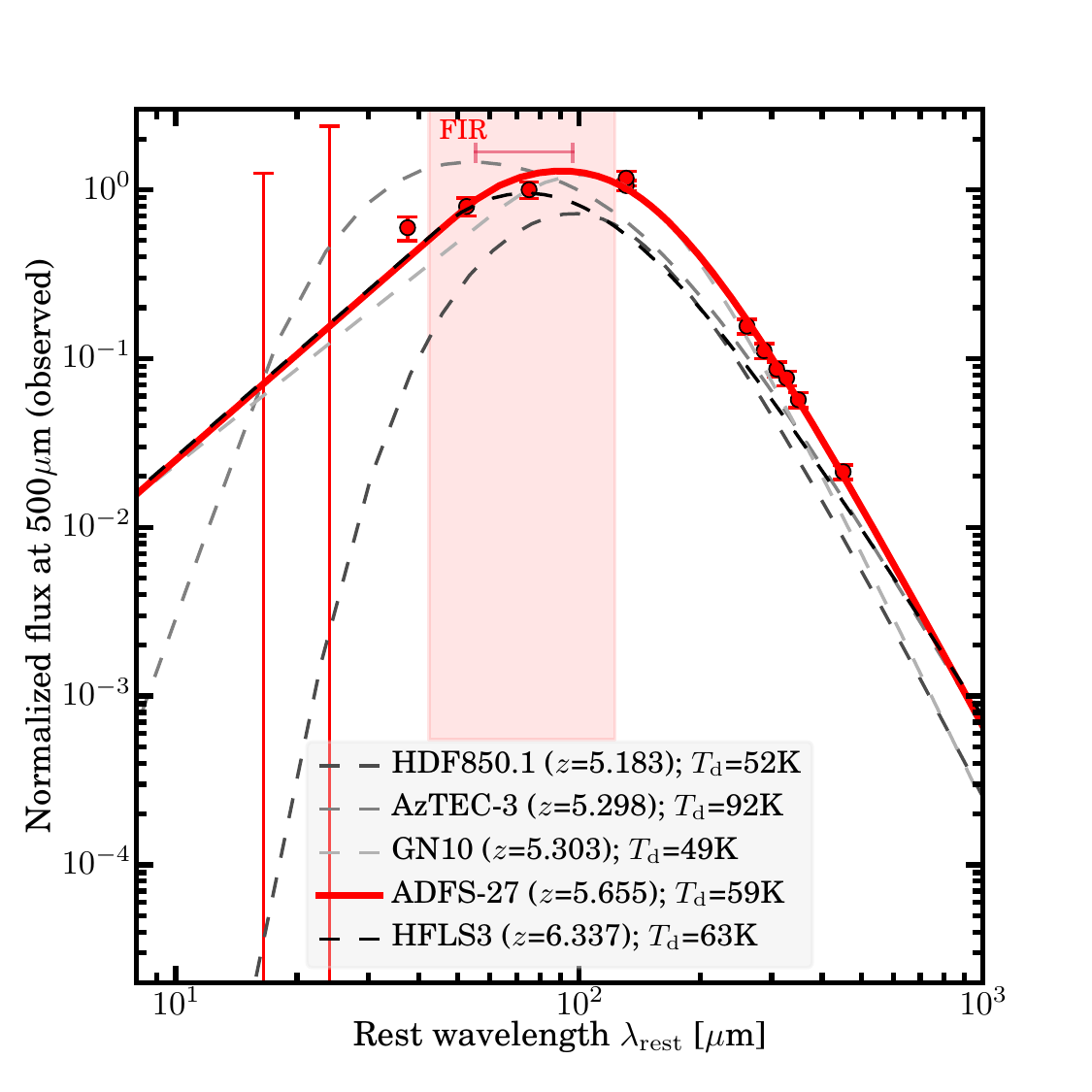}

\vspace{-4mm}

\caption{Spectral energy distribution of ADFS-27 (red line and data points), compared to Arp\,220 (left; dashed line) and four well-studied $z$$>$5 dusty starbursts from the literature (right; dashed lines; Riechers et al.\ \citeyear{riechers10a,riechers13b,riechers14b,riechers20a}; Walter et al.\ \citeyear{walter12b}), using the updated fits from Table~\ref{t3} for all sources. Left:\ {\sc CIGALE} model, also showing the individual models of ADFS-27N (light gray dotted line and points) and S (dark gray). Arp\,220 has been shifted to the observed-frame wavelength and 500\,$\mu$m flux density of ADFS-27. Right:\ All sources have been normalized to the fluxes or upper limits at observed-frame 500\,$\mu$m. The shaded region indicates the wavelength range used to calculate the far-infrared luminosities, while the entire range shown is used to calculate total infrared luminosities. The bar at the top indicates the range of peak wavelengths. \label{f7}}
\vspace{-5mm}

\end{figure*}


\begin{figure*}
\begin{deluxetable*}{ l c c c c c c c c c }

\tablecaption{Parameters obtained from dust spectral energy distribution fitting to ADFS-27 and comparison sources. \label{t3}}
\tablehead{
  Name     & redshift & $\mu_{\rm L}$\tablenotemark{a} & $T_{\rm dust}$ & $\beta_{\rm IR}$ & $\lambda_{\rm peak}$ & $\lambda_0$ & $L_{\rm FIR}$\tablenotemark{b} & $L_{\rm IR}$\tablenotemark{b} & References\\
           &          &             & (K)           &                & ($\mu$m)           & ($\mu$m)   & (10$^{12}$\,\lsol ) & (10$^{12}$\,\lsol ) & }
\startdata
HDF\,850.1 & 5.1833   & 1.6         & 51.6$^{+15.6}_{-15.1}$ & 2.67$^{+0.30}_{-0.30}$  & 85.1$^{+12.6}_{-12.6}$ & 136$^{+45}_{-49}$ & 5.2$^{+1.7}_{-1.8}$ & 7.4$^{+3.0}_{-2.9}$ & 1, 2 \\
AzTEC-3    & 5.2980   & ---         & 92.5$^{+15.4}_{-15.9}$ & 2.09$^{+0.21}_{-0.21}$  & 55.4$^{+8.4}_{-8.2}$  & 181$^{+33}_{-34}$ & 11.2$^{+1.6}_{-1.6}$ & 25.5$^{+7.3}_{-7.4}$ & 3, 4, 2 \\
GN10       & 5.3031   & ---         & 48.8$^{+9.0}_{-11.2}$  & 3.18$^{+0.26}_{-0.21}$  & 96.4$^{+9.4}_{-8.7}$  & 170$^{+20}_{-36}$ & 6.4$^{+1.1}_{-1.2}$  & 11.8$^{+1.9}_{-2.1}$ & 2 \\
ADFS-27    & 5.6550   & ---         & 59.2$^{+3.3}_{-4.1}$   & 2.52$^{+0.19}_{-0.17}$ & 85.2$^{+4.6}_{-3.9}$  & 191$^{+11}_{-19}$ & 15.8$^{+1.0}_{-1.9}$ & 23.8$^{+2.3}_{-2.2}$ & 5, 6 \\
HFLS3      & 6.3369   & 1.8$\pm$0.6 & 63.3$^{+5.4}_{-5.8}$   & 1.94$^{+0.07}_{-0.09}$  & 73.3$^{+1.6}_{-1.3}$  & 142$^{+25}_{-27}$ & 29.3$^{+1.4}_{-1.3}$ & 55.0$^{+3.0}_{-2.2}$ & 7, 2 \\
\enddata
\tablenotetext{\rm a}{Lensing magnification factor. No uncertainties are reported for HDF\,850.1 in the original work, and we assume 20\% uncertainty throughout this work.}
\tablenotetext{\rm b}{Apparent values not corrected for gravitational magnification where applicable.}
\tablereferences{[1] Walter et al.\ (\citeyear{walter12b}); [2--5] Riechers et al.\ (\citeyear{riechers20a,riechers10a,riechers14b,riechers17}); [6] this work; [7] Riechers et al.\ (\citeyear{riechers13b}).}
\vspace{-9mm}
\end{deluxetable*}

\end{figure*}


\subsubsection{Modified Black-Body Fits to the Far-Infrared Emission}

To understand the dust properties of \adfs27\ in more detail, we also
updated our modified black-body (MBB) model of the dust continuum
emission relative to that shown by Riechers et
al.\ (\citeyear{riechers17}) after including the new measurements on
the Rayleigh-Jeans tail. We adopt the same Markov Chain Monte
Carlo-based method ({\sc mbb\_emcee}) as described in our previous
work, which was initially used by Riechers et
al.\ (\citeyear{riechers13b}) and Dowell et
al.\ (\citeyear{dowell14}). The main fit parameters are the dust
temperature $T_{\rm dust}$, the spectral slope of the dust emissivity
$\beta_{\rm IR}$ and the wavelength $\lambda_0$ where the optical
depth reaches unity. We also use the observed-frame 500\,$\mu$m flux
density as a normalization parameter, and we join the MBB function to
a $\nu^\alpha$ power law toward the far Wien side of the spectral
energy distribution to better capture the decline in flux towards
shorter wavelengths, where the power-law slope $\alpha$ is used as the
fit parameter. We place a relatively broad prior of 1.8$\pm$0.6 on
$\beta_{\rm IR}$, and leave all other parameters without a prior. The
results are reported in Table~\ref{t3}, and they agree with our
previous estimates within the uncertainties. Based on the infrared
luminosity (which is comparable to the $L_{\rm dust}$ value found by
{\sc CIGALE}), we find SFR$_{\rm IR}$ =
2380$^{+230}_{-220}$\,\msol\,yr$^{-1}$ when assuming a Chabrier
(\citeyear{chabrier03}) stellar initial mass function. This is
consistent with the value reported by Riechers et
al.\ (\citeyear{riechers17}), but by about 30\% lower than the SFR
found by {\sc CIGALE}. Given the high level of dust obscuration (and
the resulting high fraction of dust-reprocessed stellar light
contributing to the total), we thus estimate that the systematic
uncertainty on the SFR and quantities derived from it is at least
30\%.

We also carried out the same fitting procedure with the same
parameters for a comparison sample of unlensed or at most
weakly-lensed $z$$>$5 dusty starbursts, namely, HDF\,850.1 ($z$=5.18),
AzTEC-3 and GN10 (both $z$=5.30), and HFLS3 ($z$=6.34; see
Table~\ref{t3}). For all sources except HDF\,850.1, this results in at
most minor changes compared to our previous work (Riechers et
al.\ \citeyear{riechers13b,riechers20a}), due to only minor differences
in the assumptions or approach. HDF\,850.1 was previously fit with a
different technique (Walter et al.\ \citeyear{walter12b}; Neri et
al.\ \citeyear{neri14}). Our new estimates include a more realistic
approach to the dust optical depth, and thus suggests a higher $T_{\rm
  dust}$, which (like the other fit parameters) still agrees with
the previous values within the uncertainties. Due to the limited
information available near the peak of the dust SED for this source,
it is expected that the detailed parameters will vary with the adopted
fitting method.

The best fits for all sources are shown in Fig.~\ref{f7}, normalized
to their observed-frame 500\,$\mu$m flux densities or estimates. The
peak wavelengths vary by a factor of 1.7, and the inferred dust
temperatures vary by a factor of 1.9. \adfs27\ falls in the middle of
the observed ranges, suggesting moderate dust properties for a $z$$>$5
dusty starburst. This is interesting, because none of the other
sources is found in an early merger stage -- which thus may not be the
dominant reason for the extreme observed properties of
\adfs27\ compared to ``typical'' galaxies at the same redshift (e.g.,
Pavesi et al.\ \citeyear{pavesi16,pavesi19}; Faisst et
al.\ \citeyear{faisst20b}). However, \adfs27\ shows the highest
$\lambda_0$, which may be the main reason for its unusually ``red''
dust SED compared to other $z$$>$5 dusty starbursts (which typically
are not ``870\,micron risers'').


\begin{figure}
\begin{deluxetable}{ l c c c c }

\tabletypesize{\scriptsize}
\tablecaption{Diameters of the gas and dust emission
in \adfs27. \label{t4}}
\tablehead{
Tracer & ADFS-27N     &       & ADFS-27S & \\
& major & minor & major & minor \\
& (mas) & (mas) & (mas) & (mas) }
\startdata
\ico\               & 415$\pm$48  & 285$\pm$56 & 344$\pm$56 & 106$\pm$101 \\
\jco\tablenotemark{a} & 464$\pm$67  & 162$\pm$89 & 417$\pm$52 & 68$\pm$126 \\
OH$^+$ (emission)   & 418$\pm$152 & 93$\pm$204 & & \\
OH$^+$ (absorption) & $<$430      & $<$120 & & \\
2.0\,mm continuum   & 401$\pm$24  & 258$\pm$22 & 422$\pm$40 & 206$\pm$49 \\
1.9\,mm continuum   & 372$\pm$21  & 239$\pm$19 & 322$\pm$24 & 138$\pm$38 \\
1.7\,mm continuum   & 356$\pm$21  & 249$\pm$32 & 371$\pm$48 & 112$\pm$93 \\
0.87\,mm continuum  & 303$\pm$30  & 213$\pm$27 & 341$\pm$31 & 146$\pm$25 \\
\enddata
\tablenotetext{\rm a}{Not corrected for contributions from the H$_2$O(3$_{12}$$\to$2$_{21}$ line).}
\end{deluxetable}
\vspace{-9mm}

\end{figure}


\subsection{Extent of the Gas and Dust Emission}

The two source components ADFS-27N and S are resolved individually in
the CO $J$=9$\to$8 and 10$\to$9 and OH$^+$ lines (and, at lower
significance, in CH$^+$ and H$_2$O) and in the 2.0\,mm to 0.87\,mm
dust continuum emission (Figs.~\ref{f3} and \ref{f6} and
Table~\ref{t4}). There is no significant difference between the extent
of the high-$J$ CO and long-wavelength dust continuum emission for
either source component. The median sizes for ADFS-27N and S from all
but the 0.87\,mm dust measurements are
(408$\pm$23)\,mas$\times$(244$\pm$27)\,mas and
(371$\pm$46)\,mas$\times$(126$\pm$14)\,mas, respectively, where the
uncertainties are the median absolute deviation. As such, any
differences between the extent of the dust at shorter and longer
wavelengths remain tentative at best, despite the finding that the
source-averaged optical depths exceed unity at observed-frame
1.3\,mm. By taking the median of all measurements, we find final sizes
of (401$\pm$29)\,mas$\times$(239$\pm$26)\,mas and
(358$\pm$26)\,mas$\times$(132$\pm$17)\,mas, for ADFS-27N and S,
respectively. This corresponds to
(2.4$\pm$0.2)\,kpc$\times$(1.4$\pm$0.2)\,kpc and
(2.2$\pm$0.2)\,kpc$\times$(0.8$\pm$0.1)\,kpc, respectively, or source
surface areas of (2.73$\pm$0.31) and (1.35$\pm$0.18)\,kpc$^2$. Based
on the updated total SFR$_{\rm IR}$ of
2380$^{+230}_{-220}$\,\msol\,yr$^{-1}$ and assuming that the flux
ratio at 0.87\,mm is representative of the peak of the SED, we find
revised SFR surface densities of $\Sigma_{\rm SFR}$=(280$\pm$110) and
(450$\pm$200)\,\msol\,yr$^{-1}$\,kpc$^{-2}$ at SFR$_{\rm IR}$ of
(1330$\pm$130) and (1050$\pm$110)\,\msol\,yr$^{-1}$ for ADFS-27N and
S, respectively. While lower than initial estimates based on fewer
data points (Riechers et al.\ \citeyear{riechers17}), this remains
consistent with the picture that these sources are both HyLIRGs and
``maximum starbursts''.\footnote{The derived $\Sigma_{\rm SFR}$ would
  be higher by a factor of 1.3 when using the SFR estimates from {\sc
    CIGALE}.}

\subsection{CO Large Velocity Gradient Modeling}

To understand the physical conditions for star formation in \adfs27,
we investigated the CO line excitation by calculating a grid of large
velocity gradient (LVG) models. Since our main goal is to obtain
constraints on the gas kinetic temperature $T_{\rm kin}$ and the H$_2$
density $\rho$(H$_2$) from collisions with CO molecules, we kept the
H$_2$ ortho-to-para ratio fixed to 3:1, the cosmic microwave background
(CMB) temperature at $z$=5.655 set to 18.135\,K, and the ratio between
CO/H$_2$ abundance and the velocity gradient fixed to
10$^{-5}$\,pc\,(\kms )$^{-1}$. We also adopted the Flower
(\citeyear{flower01}) CO collision rates (see e.g., Wei\ss\ et
al.\ \citeyear{weiss05c,weiss07}; Riechers et
al.\ \citeyear{riechers06c,riechers10a,riechers11e,riechers20a} for
similar strategies).

\begin{figure}
\epsscale{1.15}
\plotone{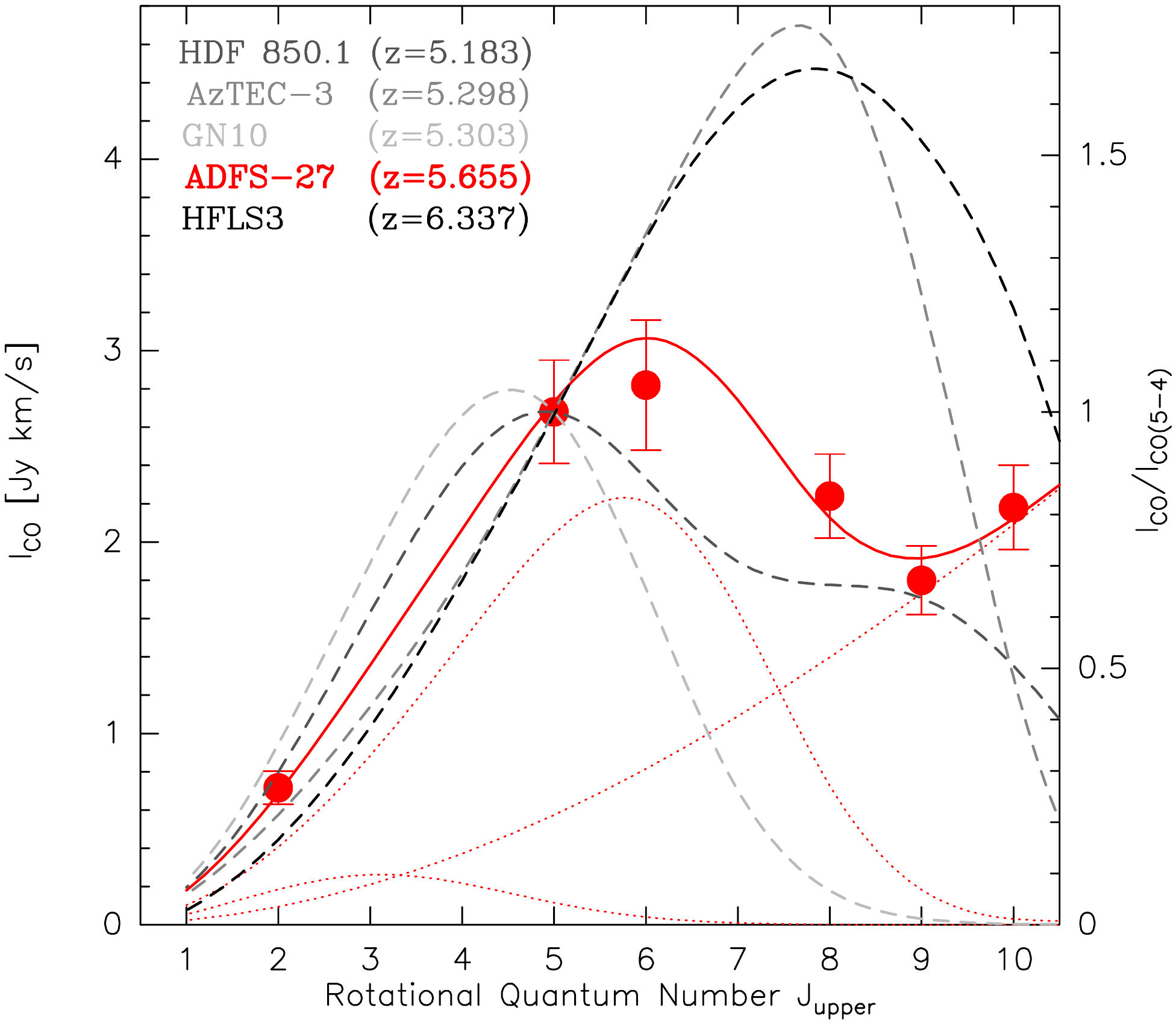}

\vspace{-2mm}

\caption{CO line ladder and LVG model of the line excitation of ADFS-27 (red lines and data points), compared to the same sources as in Fig.~\ref{f7} (dashed lines; LVG models are adopted from Riechers et al.\ \citeyear{riechers13b,riechers20a}). The dotted lines show the three LVG components for ADFS-27 with $T_{\rm kin}$=35, 40, and 100\,K and log\,[$\rho$(H$_2$)/cm$^{-3}$]=3.0, 4.0, and 6.0, respectively, while the solid line shows the sum of all components. \label{f8}}
\vspace{-4mm}

\end{figure}

The peak of the CO line ladder can be fitted well with a moderate
temperature, moderate density component, which, however, underpredicts
both the \bco\ and the \jco\ line fluxes (see Fig.~\ref{f8}). Given
the shape of the line ladder, three LVG components are required to fit
all lines simultaneously, with $T_{\rm kin}$=35, 40, and 100\,K and
log\,[$\rho$(H$_2$)/cm$^{-3}$]=3.0, 4.0, and 6.0, respectively. The
total LVG-predicted \aco\ line flux is 0.18\,Jy\,\kms, corresponding
to a CO line brightness temperature ratio between the CO $J$=2$\to$1
and 1$\to$0 lines of $r_{21}$=0.95. As such, the measured \bco\ line
profile is expected to be a good representation of the total CO
emission. From our model, we find that 57\% of the \aco\ emission are
associated with the gas component that dominantly contributes at the
peak of the line ladder, with only 30\% being associated with the
lowest-excitation component (which, however, may dominate the total
gas mass). Only 13\% is due to the highest-excitation component, which
is required to reproduce the observed CO $J$=10$\to$9 to 9$\to$8
ratio, such that it is expected to contribute relatively little to the
total gas mass (or dust mass). This component only becomes a dominant
contributor to the CO line luminosity in the \hco\ line and
above. Overall, we are able to fully explain the CO excitation up to
the $J$=10$\to$9 level with collisional excitation, without requiring
unusually high $T_{\rm kin}$ or $\rho$(H$_2$) compared to the
conditions that exist in molecular cloud cores. We also do not find
evidence for very cold gas, but such emission would be challenging to
detect given the CMB temperature of close to 20\,K at this
redshift. We find that the line flux ratio between ADFS-27N and S
increases by only 4\% between the CO $J$=9$\to$8 and 10$\to$9 lines,
which is insignificant compared to the relative calibration
uncertainties. As such, further measurements are required to
investigate any potential difference in CO excitation between the two
merging galaxies.

We compare the CO line excitation in \adfs27\ to those found in the
same sample studied above in the dust continuum analysis (dashed lines
in Fig.~\ref{f8}; models adopted from Riechers et
al.\ \citeyear{riechers13b,riechers20a}). We find that \adfs27\ has a
moderate CO excitation for a $z$=5 dusty starburst galaxy, but we
caution that the only other source for which measurements up to the CO
$J$=10$\to$9 transition have been reported is HFLS3 at $z$=6.34. As
such, the presence of potentially warmer, high-excitation components
cannot be ruled out for HDF\,850.1 and GN10, i.e., the sources with
apparently lower CO excitation than \adfs27. However, we note that
these are also the two sources with the lowest $T_{\rm dust}$ levels,
such that the level of CO excitation, to first order, appears to
parallel that seen in the dust continuum SED shapes, with the warmest
sources showing the highest CO excitation. Moreover, to first order,
this parallels the trend in increasing CO excitation with increasing SFR
surface density previously proposed by Riechers et
al.\ (\citeyear{riechers20a}).

\subsection{Dynamical Masses, Gas Mass Fraction, and Gas-to-Dust Ratio}

The CO emission in ADFS-27N and S is resolved both spatially and in
velocity space, showing similar gradients in the CO $J$=9$\to$8 and
10$\to$9 lines (Fig.~\ref{f4}). Since the patterns are comparable, but
the \jco\ line is difficult to de-blend from the H$_2$O contribution
to the velocity field, we focus on the \ico\ kinematic structure in
the following. From blue to red, the centroid of the CO emission in
ADFS-27N moves by approximately 0.25$''$ (1.5\,kpc) from north-west to
south-east, where the east-west extension marks the dominant
contribution (0.24$''$) relative to north-south (maximum of
0.14$''$). The CO centroid in ADFS-27S moves by 0.19$''$ (1.2\,kpc),
and almost entirely in the north-south direction (0.19$''$ versus
maximum east-west motion of 0.02$''$). The source sizes are consistent
with point sources in all but their central velocity channels, showing
that the physical extent of the galaxies in their moment-0 maps is
mostly due to spatially-resolved velocity gradients. In their central
channels, ADFS-27N has a \ico\ diameter of
(517$\pm$67)\,mas$\times$(236$\pm$71)\,mas, and ADFS-27S is
(328$\pm$63)\,mas$\times$(101$\pm$74)\,mas. These are comparable to
the sizes in the moment-0 maps, but ADFS-27N is consistent with being
$\sim$30\% more extended. The velocity gradients are consistent with
disk-like rotation in both cases, but higher spatial resolution is
required to better distinguish between rotation- and
dispersion-dominated components since constraints on the CO velocity
dispersion are currently largely limited to the galaxy-wide
dispersion.

To obtain dynamical mass estimates, we adopt an isotropic virial estimator
(e.g., Engel et al.\ \citeyear{engel10}), and sizes that are 30\%
larger than the averages obtained from the moment-0 and continuum maps
for ADFS-27N to account for the findings in the previous
paragraph. For the \ico\ line widths, this yields $M_{\rm dyn}^{\rm
  N}$=(4.4$\pm$0.3)$\times$10$^{11}$\,\msol, and $M_{\rm dyn}^{\rm
  S}$=(3.6$\pm$0.3)$\times$10$^{11}$\,\msol, for a total value of $M_{\rm
  dyn}$=(8.0$\pm$0.4)$\times$10$^{11}$\,\msol\ across both components.  Using
the stellar and dust masses found in our analysis,
  and neglecting any potential mass contributions from dark matter, we
  find upper limits on the $\alpha_{\rm CO}$ conversion factor from CO
  luminosity to molecular gas mass of $<$2.4 and
  3.3\,\msol\,(K\,\kms\,pc$^{-2}$)$^{-1}$ for ADFS-27N and S,
  respectively. If we assume a 25\% contribution to $M_{\rm dyn}$ due
  to dark matter (e.g., Daddi et al.\ \citeyear{daddi10a}), these
  limits are reduced to $\alpha_{\rm CO}^{\rm dyn}$$<$1.5 and
  2.3\,\msol\,(K\,\kms\,pc$^{-2}$)$^{-1}$. This suggests that an
  assumption of $\alpha_{\rm
    CO}$=1.0\,\msol\,(K\,\kms\,pc$^{-2}$)$^{-1}$ is reasonable for a
  merger-driven starburst at $z$$\sim$6 like \adfs27. We thus adopt
  this value throughout this work.  This corresponds to gas mass
  fractions of $f_{\rm gas}$=0.26 for each source component.
When assuming the \ico\ line flux ratio as the gas mass ratio, we find
gas-to-dust ratios of 45$\pm$8 and 65$\pm$10 for ADFS-27N and S,
respectively, or a source-averaged value of 50$\pm$6. This is
comparable to what is found for the $z$=6.34 dusty starburst HFLS3
(Riechers et al.\ \citeyear{riechers13b}), and compatible with the
values found for nearby star-forming galaxies (e.g., Wilson et
al.\ \citeyear{wilson08}). It also is compatible with the earlier
value found by Riechers et al.\ (\citeyear{riechers17}) under common
assumptions.


\begin{figure*}
\begin{deluxetable}{ l c c c c }

\tablecaption{Derived quantities for \adfs27. \label{t5}}
\tablehead{
Quantity & Unit & ADFS-27 & ADFS-27N & ADFS-27S }
\startdata
$M_{\rm gas}$  & 10$^{11}$\,\msol & 2.1$\pm$0.2 & 1.16$\pm$0.10 & 0.94$\pm$0.08 \\
$M_{\rm dust}$ & 10$^{9}$\,\msol & 4.2$\pm$0.4 & 2.6$\pm$0.4 & 1.5$\pm$0.2 \\
$M_{\star}$    & 10$^{11}$\,\msol & 2.1$\pm$0.6 & 1.6$\pm$1.3\tablenotemark{a} & 0.46$\pm$0.15 \\
$M_{\rm dyn}$  & 10$^{11}$\,\msol & 8.0$\pm$0.4 & 4.4$\pm$0.3 & 3.6$\pm$0.3 \\
SFR$_{\rm IR}$ & \msol\,yr$^{-1}$ & 2380$^{+230}_{-220}$ & 1330$\pm$130 & 1050$\pm$110 \\
$t_{\rm age}$  & Myr             & & 110$\pm$60 & 80$\pm$30 \\
GDR=$M_{\rm gas}$/$M_{\rm dust}$ &  & 50$\pm$6 & 45$\pm$8 & 65$\pm$10 \\
DY[$t_{\rm age}$]=$M_{\rm dust}$/$M_{\star}$ &  & 2\% & 1.6\% & 3.2\% \\
$f_{\rm gas}$=$M_{\rm gas}$/$M_{\rm dyn}$ & & 0.26$\pm$0.02 & 0.26$\pm$0.03 & 0.26$\pm$0.03 \\
$\alpha_{\rm CO}^{\rm dyn}$ & \msol\,(K\kms\,pc$^2$)$^{-1}$ & & $<$1.5 & $<$2.3 \\
$t_{\rm dep}$=$M_{\rm gas}$/SFR\tablenotemark{b} & Myr & 88$\pm$10 & 87$\pm$12 & 89$\pm$12 \\
$\Sigma_{\rm gas}$ & 10$^{4}$\,\msol\,pc$^2$ & $\leq$3.0 & $\leq$1.9 & $\leq$4.0 \\
$\Sigma_{\rm SFR}$\tablenotemark{b} & \msol\,kpc$^2$ & 340$\pm$140 & 280$\pm$110 & 450$\pm$200 \\
\enddata
\tablenotetext{\rm a}{We estimate this value to have a systematic uncertainty of a factor of three.}
\tablenotetext{\rm b}{Values would be 1.3 times lower/higher when assuming the SFR of 3100$\pm$330\,\msol\,yr$^{-1}$ found by {\sc CIGALE}, which we consider to be the systematic uncertainty for these quantities.}
\end{deluxetable}
\vspace{-9mm}

\end{figure*}



\subsection{Gas Mass, Surface Density, and Depletion Time}

Based on the LVG-predicted \aco\ line luminosity, we find a total cold
molecular gas mass of $M_{\rm
  gas}$=(2.1$\pm$0.2)$\times$10$^{11}$\,$(\alpha_{\rm
  CO}/1.0)$\,\msol, where $\alpha_{\rm CO}$ is the conversion factor
from CO luminosity to molecular gas mass in units of
\msol\,(K\,\kms\,pc$^{-2}$)$^{-1}$. Based on the extent of the
high-$J$ CO and dust emission, this provides an upper limit on the gas
surface density of $\Sigma_{\rm
  gas}$$\leq$3.0$\times$10$^4$\,\msol\,pc$^{-2}$. When assuming the
\ico\ line flux ratio, this also suggests $\Sigma_{\rm gas}^{\rm
  N}$$\leq$1.9$\times$10$^4$\,\msol\,pc$^{-2}$ and $\Sigma_{\rm
  gas}^{\rm S}$$\leq$4.0$\times$10$^4$\,\msol\,pc$^{-2}$ for ADFS-27N
and S, respectively. This is approximately consistent with the
star-formation law found at lower redshift (e.g., Hodge et
al.\ \citeyear{hodge15}). It also corresponds to an overall gas
depletion timescale of $t_{\rm dep}$$\sim$90$\pm$10\,Myr, which is
consistent with expectations for starburst galaxies (e.g., Carilli \&
Walter \citeyear{cw13}).\footnote{The derived $t_{\rm dep}$ would be
  by a factor of 1.3 shorter when using the SFR estimates from {\sc
    CIGALE}, but this could be compensated by a higher $\alpha_{\rm
    CO}$, as allowed by the data within the uncertainties.}

\subsection{OH$^+$ and CH$^+$ Optical Depths and Column Densities}

Together with HFLS3 ($z$=6.34; see Fig.~\ref{f6b} and Riechers et
al.\ \citeyear{riechers13b}), ADFS-27N is only the second $z$$\sim$6
galaxy that shows absorption/emission from OH$^+$ and CH$^+$. The (on
average) blueshifted OH$^+$ and CH$^+$ absorption toward ADFS-27N is
expected to come from a cool ($\lesssim$100\,K), low-density
($<$100\,cm$^{-3}$), spatially-extended gas component, while the
broad, redshifted OH$^+$ emission component is likely associated with
a shock-heated, dense gas component impacted by galactic winds (e.g.,
Falgarone et al.\ \citeyear{falgarone17}). The OH$^+$ emission thus
may be associated with the densest gas component in the ``maximum
starburst'' nucleus that leads to the enhanced \jco\ emission, and the
spatial and redshift offset may suggest that it is associated with a
wind emerging from this region. The optical depth of an unsaturated
absorption line is $\tau_{\rm line}$=--ln($f_{\rm trans}$), where
$f_{\rm trans}$ is the fraction of the continuum emission that is
still transmitted. Averaged over the entire line widths, we find
$\tau_{\rm OH+}$=0.35$\pm$0.04 for the absorption component of the
OH$^+$(1$_1$$\to$0$_1$) line and $\tau_{\rm CH+}$=0.38$\pm$0.18 for
the tentative CH$^+$($J$=1$\to$0) line observed toward ADFS-27N. This
corresponds to velocity-integrated optical depths of 234$\pm$26 and
415$\pm$198\,\kms, respectively. The difference between both molecules
is not significant, and largely due to the broader velocity range
adopted for the tentative CH$^+$ feature. As such, we conclude that
the OH$^+$ and CH$^+$ optical depths are indistinguishable at the
quality level of our current data. The CH$^+$ peak optical depth is
comparable to those found for strongly-lensed dusty starbursts at
$z$$\sim$2 (Falgarone et al.\ \citeyear{falgarone17}).

Following Equation 5 of Indriolo et al.\ (\citeyear{indriolo18}) and
Falgarone et al.\ (\citeyear{falgarone17}), we find OH$^+$ and CH$^+$
column densities of (11.4$\pm$1.2) and
(12.5$\pm$5.9)$\times$10$^{14}$\,cm$^{-2}$, which is comparable to
what is found for the Cosmic Eyelash at $z$=2.3 by the same authors,
despite a roughly 5 times higher star-formation rate in ADFS-27N. These
authors suggest that estimates based on this ground-state OH$^+$
transition should be a good representation of the total OH$^+$ column
density, given that the CMB temperature even at $z$=5.7 is
significantly below that necessary to populate the first rotationally
excited state of OH$^+$. Based on the modeling carried out by Indriolo
et al., this is consistent with a cosmic ray ionization rate of
$\zeta_{\rm H}$$\sim$10$^{-15}$\,s$^{-1}$ in the gas found in a
low-density, extended gaseous halo around ADFS-27N that gives rise to
the OH$^+$ and CH$^+$ absorption. In this regard, the non-detection of
absorption towards ADFS-27S is interesting, since it may suggest that
the diffuse gaseous halo does not extend out to or beyond 9\,kpc from
the central source, despite the early-stage merger between ADFS-27N and
S. On the other hand, the redshift difference of
(140$\pm$13)\,\kms\ may indicate that ADFS-27S is in the foreground of
ADFS-27N, such that the de-projected distance may exceed 10\,kpc or
more.

Following Falgarone et al.\ (\citeyear{falgarone17}), we estimate that
the hydrogen mass in the diffuse halo is 5\% of the total gas mass of
\adfs27, which we consider to be uncertain by a factor of a few. The
absolute value of 0.6$\times$10$^{10}$\,\msol\ is comparable to those
found for strongly-lensed $z$$\sim$2 dusty starbursts (Falgarone et
al.\ \citeyear{falgarone17}). To create such a mass reservoir on
45\,Myr timescales (i.e., approximately half the gas depletion time)
through gaseous outflows would require an mass outflow rate of
125\,\msol\,yr$^{-1}$, corresponding to 9\% of the current SFR of
ADFS-27N. Thus, it appears realistic to assume that the extended gas
reservoir is maintained by a starburst-driven outflow from the
``maximum starburst'' nucleus. Given the lack of a detection of any
OH$^+$ or CH$^+$ absorption toward the more compact,
higher-$\Sigma_{\rm SFR}$ component ADFS-27S, we conclude that such
diffuse gas reservoirs in the environment of the most intense
starbursts may not be entirely spherical. Since ADFS-27N appears to be
the older source component by 30\,Myr, it could also be possible that
enhanced OH$^+$ features preferentially appear in slightly later
stages of the evolution of HyLIRGs. However, the ages of both source
components agree within the uncertainties, such that this conclusion
remains speculative.

\subsection{On the ``870\,micron riser'' population}

Since the initial discovery of \adfs27, one additional ``870\,micron
riser'' has been reported, SPT\,0245--63 at $z$=5.6256 (Reuter et
al.\ \citeyear{reuter20}). The same work has also revised the redshift
of the other previously known example, SPT\,0243--49, to
$z$=5.7022. Together with the redshifts of ADFS-27N and S, this yields
an average redshift of $z$=5.659 for the currently known population,
with a surprisingly narrow range in redshift of $\pm$0.038 that may be
dominantly due to the currently small sample size.

\section{Conclusions}

We have detected CO $J$=2$\to$1 to 10$\to$9 emission toward the
$z$=5.7 binary HyLIRG \adfs27, revealing a massive, moderately highly
excited molecular gas reservoir with a mass of $M_{\rm
  gas}$=(2.1$\pm$0.2)$\times$10$^{11}$\,$(\alpha_{\rm
  CO}/1.0)$\,\msol, corresponding to about 25\% of its stellar
mass. The two starbursts are separated by only 140\,\kms\ along the
line of sight, and 9.0\,kpc in projection. The kinematic structure
seen in high-$J$ CO line emission is consistent with 2.3\,kpc diameter
rotating disks in both cases, but the factor of about 2 difference in
minor axis length may suggest differences in orientation toward the
line of sight, unless the sources are intrinsically asymmetric. The
dynamical mass estimates suggest that ADFS-27 is consistent with a 1:1
merger of very massive galaxies that have comparable gas fractions,
which at face value would suggest a similar evolutionary stage. A
detection of high-level ($J_{\rm upper}$=3) H$_2$O lines and
ground-state OH$^+$ transitions in emission, together with an
apparently enhanced \jco\ flux, reveals the presence of a high
density, high kinetic temperature gas component and a strong infrared
radiation field embedded in the gas reservoir, which is likely
associated with the cores of the ``maximum starburst'' nuclei. The
H$_2$O emission appears to be significantly stronger in the southern
starburst ADFS-27S, which may indicate a higher radiation field
intensity. On the other hand, a blueshifted absorption component is
seen in the OH$^+$ and (tentatively) CH$^+$ ground-state transitions,
together with redshifted OH$^+$ emission, in the northern starburst
ADFS-27N, whereas neither of these molecular species is detected in
ADFS-27S. This suggests the presence of a massive molecular outflow
from the northern source, which feeds enriched gas to its halo at a
rate of approximately 125\,\msol\,yr$^{-1}$, or 9\% of its SFR. The
lack of OH$^+$ and CH$^+$ absorption in ADFS-27S could be understood
if the distribution of the diffuse gas has a relatively low covering
fraction. However, this interpretation is not preferred when taking
into account the finding that CH$^+$ absorption in particular appears
to be ubiquitous towards $z$$\sim$2 lensed starbursts (Falgarone et
al.\ \citeyear{falgarone17}).

ADFS-27S appears to be the warmer source, consistent with its stronger
H$_2$O emission, but the dust temperature of $T_{\rm dust}$=59\,K
averaged over both source components is moderate for a $z$$>$5 dusty
starburst (which have a median $T_{\rm dust}$=50\,K; Riechers et
al.\ \citeyear{riechers20a}). A difference in dust temperature between
the two starbursts would be consistent with the unusual shape of the
peak of its dust SED, which has currently not been resolved into the
two components on the Wien side. Stellar light is detected toward
ADFS-27S in the rest-frame ultraviolet, while ADFS-27N remains
undetected below at least observed-frame 3.6\,$\mu$m. Modeling of the
full spectral energy distribution suggests that the deeply-embedded,
``optically-dark'' source component ADFS-27N is the older of the two
sources (albeit consistent within the uncertainties), with an age of
only 110\,Myr (compared to 80\,Myr for ADFS-27S). However, it cannot
be ruled out that an older stellar population is completely hidden
from our view in ADFS-27N, and it would likely be outshone by the
light from young massive stars in ADFS-27S. The inferred ages of the
stellar populations are comparable to the gas depletion times in both
sources (which themselves are comparable to other $z$$>$5 dusty
starbursts, e.g., Riechers et al.\ \citeyear{riechers20a}), which may
suggest an intermediate stage of their starbursts, despite the
relatively early stage of the merger.

Overall, the two merging HyLIRGs that are part of the ADFS-27 system
show decisively different physical properties, several of which cannot
be solely explained by orientation or line-of-sight differences. This
may suggest at least some differences in their evolutionary stages, or
in the gas supply that powers their starbursts and the resulting
feedback. The extended diffuse gas reservoir associated with ADFS-27N
seen in OH$^+$ and CH$^+$ in absorption can plausibly be put in place
through feedback on tens of millions of years timescales and supported
by its current gas reservoir, such that even a 30\,Myr difference in
the onset times of the starbursts could explain a fair range of the
differences seen between both sources. As such, our findings on the
feedback timescales are consistent with those found for
strongly-lensed $z$$\sim$2 starbursts (e.g., Falgarone et
al.\ \citeyear{falgarone17}).

After concluding the major merger process and depleting its large gas
reservoir through intense star formation, ADFS-27 will likely evolve
into one of the most massive galaxies by $z$=3--4, with a stellar mass
of order 4$\times$10$^{11}$\,\msol\ or more (depending on additional
gas supply). While major mergers may not be the dominant process
leading to galaxy mass assembly in the universe, this event will be
important to form such a rare, massive galaxy early in the universe's
history. Despite its relatively short gas depletion timescale of
90\,Myr, at least part of the gas mass will be converted into stars on
longer timescales, due to the already observed feedback processes,
extending the star formation period to likely at least a few hundred
million years, even before accounting for additional gas accretion from
larger scales from the environment of ADFS-27.

Given the lack of identifications of other, similar 1:1 mergers of a
few times 10$^{11}$\,\msol\ galaxies at comparable
redshifts,\footnote{Other, at least somewhat similar examples are
  known at $z$=2--4, e.g., Fu et al.\ (\citeyear{fu13}); Emonts et
  al.\ (\citeyear{emonts15}); Oteo et al.\ (\citeyear{oteo16}).} a
detailed study of the ADFS-27 system is of particular interest to
understand such a rare, relatively short-lived phase in the early
evolution of massive galaxies. Higher-resolution ALMA data will be
capable of revealing the sub-kiloparsec scale gas dynamics that are
not yet accessible with the current data, but which are required to
reveal if the two merger components truly are rotating, smooth disks,
or if they already show substantial pertubation of their velocity
fields. ALMA will also be able to better constrain the properties of
the extended diffuse gas component, setting the stage for similar
investigations in low-$J$ CO emission with the Next Generation Very
Large Array (ngVLA; e.g., Murphy et al.\ \citeyear{murphy18}) for
galaxies in the northern sky in the coming decade. At the same time,
{\em JWST} will be able to better constrain the properties of the
stellar populations, in particular for ADFS-27N, which remains
undetected in our deep {\em HST} imaging due to dust
obscuration. Given the extreme rarity of systems like ADFS-27, a more
precise measurement of its stellar mass and age will be key to
constrain the parameter space that needs to be explored by simulations
of galaxy formation to account for the enigmatic HyLIRG population
that already existed within the first billion years of cosmic time.

\acknowledgments

The authors thank the anonymous referee for helpful comments that have
resulted in improvements in the manuscript. The authors also thank
Ivan Oteo and Ariadna Manilla Robles for help with carrying out the
ATCA observations, and Rychard Bouwens for help with fine-tuning the
selection criteria for Lyman-break galaxies for the redshift of
\adfs27.  D.R. acknowledges support from the National Science
Foundation under grant numbers AST-1614213 and AST-1910107, and from
NASA through grant number HST-GO-15919. D.R. also acknowledges support
from the Alexander von Humboldt Foundation through a Humboldt Research
Fellowship for Experienced Researchers.  H.N. and A.C. acknowledge
support from NASA 80NSSC20K0437 and HST-GO-15919. D.Rig. acknowledges
support from STFC through grant ST$/$S000488$/$1 and the University of
Oxford John Fell Fund. The National Radio Astronomy Observatory is a
facility of the National Science Foundation operated under cooperative
agreement by Associated Universities, Inc. This paper makes use of the
following ALMA data: ADS/JAO.ALMA\#\,2017.1.00235.S; 2018.1.00966.S;
2016.1.00613.S; and 2013.1.00001.S. ALMA is a partnership of ESO
(representing its member states), NSF (USA) and NINS (Japan), together
with NRC (Canada) and NSC and ASIAA (Taiwan), in cooperation with the
Republic of Chile. The Joint ALMA Observatory is operated by ESO, AUI/
NRAO and NAOJ. The Australia Telescope Compact Array is part of the
Australia Telescope National Facility, which is funded by the
Australian Government for operation as a National Facility managed by
CSIRO. This research is based on observations made with the NASA/ESA
{\em Hubble Space Telescope} obtained from the Space Telescope Science
Institute, which is operated by the Association of Universities for
Research in Astronomy, Inc., under NASA contract NAS 5-26555. These
observations are associated with programs 15919 and 15464.

\vspace{5mm}
\facilities{ALMA, ATCA, Hubble(ACS and WFC3)}

 \software{MIRIAD (Sault, Teuben \& Wright \citeyear{sault95}), CASA package (v5.6.1; McMullin et al.\ \citeyear{mcmullin07}), CIGALE (Burgarella et al.\ \citeyear{Burgarella2005}; Noll et
al.\ \citeyear{Noll2009}; Boquien et al.\ \citeyear{Boquien2019})}

\bibliographystyle{yahapj}
\bibliography{ref.bib}


\end{document}